\documentclass[pdflatex,sn-nature]{sn-jnl}

\usepackage{lineno}
\usepackage{caption}
\usepackage{graphicx}%
\usepackage{multirow}%
\usepackage{amsmath,amssymb,amsfonts}%
\usepackage{amsthm}%
\usepackage{mathrsfs}%
\usepackage[title]{appendix}%
\usepackage{xcolor}%
\usepackage{textcomp}%
\usepackage{hyperref}
\usepackage{manyfoot}%
\usepackage{booktabs}%
\usepackage{algorithm}%
\usepackage{algorithmicx}%
\usepackage{algpseudocode}%
\usepackage{listings}%
\usepackage[utf8]{inputenc}
\usepackage{graphicx}
\theoremstyle{thmstyleone}%
\usepackage{soul}
\theoremstyle{thmstyletwo}%

\theoremstyle{thmstylethree}%

\makeatletter
\renewcommand\@biblabel[1]{#1.}
\makeatother

\begin{document}

\title[Article Title]{Similar evapotranspiration responses mask contrasting modes of water access during drought in California}

\author*[1]{\fnm{Yan} \sur{Zhang}}\email{yanz74@uci.edu}

\author[2]{\fnm{Veronica J.} \sur{Berrocal}}

\author[3]{\fnm{Joshua B.} \sur{Fisher}}

\author[4]{\fnm{Octavia} \sur{Crompton}}

\author[5]{\fnm{Alvar} \sur{Escriva-Bou}}

\author[1]{\fnm{Angela J.} \sur{Rigden}}

\affil*[1]{\orgdiv{Department of Earth System Science}, \orgname{University of California}, \orgaddress{\city{Irvine}, \state{CA}, \country{USA}}}

\affil[2]{\orgdiv{Department of Statistics}, \orgname{University of California}, \orgaddress{\city{Irvine}, \state{CA}, \country{USA}}}

\affil[3]{\orgdiv{Schmid College of Science and Technology}, \orgname{Chapman University}, \orgaddress{\city{Orange}, \state{CA}, \country{USA}}}

\affil[4]{\orgdiv{Agricultural Research Service Hydrology and Remote Sensing Laboratory}, \orgname{USDA}, \orgaddress{\city{Beltsville}, \state{MD}, \country{USA}}}

\affil[5]{\orgdiv{Department of Civil and Environmental Engineering}, \orgname{University of California}, \orgaddress{\city{Davis}, \state{CA}, \country{USA}}}

\abstract{Atmospheric warming raises evaporative demand, but actual evapotranspiration (ETa) may track demand or decouple under water limitation. Here, we quantified demand-tracking ability from regression slopes relating deseasonalized ETa and reference evapotranspiration (ETo) anomalies across California during 2008–2024. Forests and croplands retained similar positive tracking under severe drought, whereas grasslands and shrublands showed weak or negative responses. Despite similar demand tracking, forest responses were consistent with subsurface-storage access, whereas cropland responses reflected managed water access and crop-system adjustment. Positive ETa–ETo coupling on active fields coexisted with a higher statewide crop-attributed fallow share of managed cropland than under no drought, showing that active-field estimates alone did not capture whole-crop-system adjustment. Late-period demand-tracking ability was lower at 54.9\% of grid cells, with larger regional declines accompanying greater groundwater-storage losses. Our results indicate that similar ETa responses mask contrasting modes of water access, requiring joint consideration of hydrology and management when interpreting drought responses.}

\maketitle

\section*{Main}

Atmospheric warming is reframing terrestrial water scarcity as a mismatch between atmospheric evaporative demand and accessible water, rather than as a precipitation deficit alone \citep{bergLandAtmosphereFeedbacks2016, denissenWidespreadShiftEcosystem2022}. Rising temperature and vapor pressure deficit (VPD) increase the atmospheric evaporative demand for water from land surfaces, while water availability varies unevenly across space and time \citep{gebrechorkosWarmingAcceleratesGlobal2025, novickImpactsRisingVapour2024, seneviratneInvestigatingSoilMoisture2010}. The same demand anomaly can therefore produce contrasting land surface responses: where water remains accessible, evapotranspiration can continue to increase with demand, whereas under water limitation high demand can coincide with reduced evapotranspiration. Land-surface feedbacks can further amplify drying as declining soil moisture reduces evaporative cooling and increases atmospheric aridity \citep{seneviratneInvestigatingSoilMoisture2010, bergLandAtmosphereFeedbacks2016}. The hydrological consequences of rising demand therefore depend on whether actual evapotranspiration (ETa) remains coupled to atmospheric evaporative demand during drought, or decouples as water availability becomes limiting \citep{budykoClimateLife1974, novickIncreasingImportanceAtmospheric2016, jungRecentDeclineGlobal2010}.

Reference ET (ETo) provides a standardized measure of atmospheric evaporative demand for a well-watered reference surface \cite{allenrichardg.CropEvapotranspirationGuidelines1998, fisherComeHomePotential2011}. Actual evapotranspiration (ETa), in contrast, represents the land-to-atmosphere water flux under prevailing water availability, vegetation regulation, and management. The relationship between ETo and ETa therefore provides a common diagnostic of whether accessible water is sufficient to sustain ETa as atmospheric evaporative demand rises during drought. Positive coupling indicates that higher ETo is accompanied by higher ETa, whereas weak or reversed coupling indicates increasing constraints from soil moisture depletion, stomatal regulation, or reduced canopy conductance \cite{novickIncreasingImportanceAtmospheric2016, grossiordPlantResponsesRising2020, seneviratneInvestigatingSoilMoisture2010, jungRecentDeclineGlobal2010}. Complementary evaporation theory further implies that reduced ETa can warm and dry the near-surface atmosphere, increasing potential or reference evaporation \cite{BouchetRJ1963a, brutsaertAdvectionaridityApproachEstimate1979, fisherGlobalEstimatesLand2008}. The key signal is therefore not atmospheric evaporative demand alone, but whether ETa remains coupled to atmospheric evaporative demand under water stress.

Previous work has shown that drought can limit ETa through different mechanisms in natural and managed systems. In natural ecosystems, drought responses are interpreted through rooting depth, hydraulic traits, stomatal regulation, xylem vulnerability, and access to stored or bedrock water \cite{mcdowellDarcysLawPredicts2015, andereggHydraulicDiversityForests2018, mccormickWidespreadWoodyPlant2021}. In agricultural systems, drought responses are interpreted in terms of crop coefficients, irrigation demand, water stress, and satellite-based ETa estimates over managed fields \cite{allenrichardg.CropEvapotranspirationGuidelines1998, volkAssessingAccuracyOpenET2024, boserFieldscaleCropWater2024}. Thus, similar ETa responses to atmospheric evaporative demand during drought do not necessarily imply similar modes of water access. Comparing demand responses across natural and agricultural systems therefore requires both a common demand response metric and land-system context to interpret the processes sustaining that response. Natural systems provide a hydrologic reference because their responses are not sustained by irrigation or crop-allocation decisions. Cropland tracking can then be interpreted relative to that reference, helping distinguish managed water access from background hydrologic conditions.

California provides a strong setting for this comparison. The state spans forests, grasslands, shrublands and extensive croplands across strong ecological and hydroclimatic gradients (Supplementary Fig. S1a). It also combines rising growing-season atmospheric evaporative demand, shown by increasing VPD and ETo in the California gridMET record (Supplementary Fig. S1b,c; \cite{Abatzoglou2013}), with repeated drought exposure during 2008-2024 (Supplementary Fig. S1d; \cite{SvobodaEtAl2002, williamsRapidIntensificationEmerging2022}). Within this landscape, Central Valley irrigation sustains one of the world's most productive agricultural regions, whose output reaches consumers across the United States and internationally \cite{marstonDroughtImpactsWater2017}. Evapotranspiration returns part of the water applied to these croplands to the atmosphere, altering local heat and moisture exchange and potentially influencing moisture transport and summer precipitation beyond the valley \cite{yangImpactIrrigationCalifornia2017,mcdermidIrrigationEarthSystem2023}. During drought, groundwater can provide two-thirds or more of this irrigation water in the Central Valley, and GRACE-based analyses show accelerating depletion during the 2019–2021 drought phase of the 2000–2021 southwestern North American megadrought \cite{liuGroundwaterDepletionCalifornias2022}. More broadly, rapid groundwater declines are widespread in dry cultivated regions \cite{jasechkoRapidGroundwaterDecline2024}, and aquifer depletion can reduce irrigated agriculture's ability to buffer production against water deficits \cite{mienoAquiferDepletionExacerbates2024}. California therefore combines ecological water access, irrigation management, land-use adjustment, and groundwater-storage stress within one natural and managed landscape system.

Here, we quantify demand-tracking ability across California's natural ecosystems and managed croplands from 2008 to 2024. We define demand-tracking ability as the slope $\beta$ between deseasonalized growing-season anomalies of ETa and ETo, so that the metric captures anomaly-scale coupling rather than their shared seasonal cycle. We investigate whether similar ETa responses to atmospheric evaporative demand carry the same meaning across forests, dryland ecosystems and irrigated croplands, and whether crop-system accounting, which includes fallow as well as active fields, changes the interpretation of cropland tracking. By pairing demand tracking with land-system context, we then distinguish five severe-drought response states: positive natural-system coupling, where natural systems sustain ETa as demand rises, consistent with access to subsurface storage; water-limited decoupling, where they do not; managed tracking, where the whole crop system sustains ETa with little change in fallow; managed contraction, where whole-system tracking remains positive but the active-field versus whole-system tracking gap or the increase in fallow share is large; and weak managed response, where the crop system as a whole does not track demand.

\subsection*{Evapotranspiration responses to atmospheric evaporative demand diverge across California land systems}

We first mapped demand-tracking ability to identify where unusually high atmospheric evaporative demand was matched by higher ETa after removing the seasonal cycle. Demand-tracking ability was estimated from deseasonalized monthly growing-season ETa and ETo anomalies. Positive values indicate that ETa increased when ETo was anomalously high, a pattern consistent with water remaining sufficiently accessible to support higher ETa. Values near zero indicate little ETa response, whereas negative values indicate that ETa declined as ETo increased. In either case, ETa did not follow increases in atmospheric evaporative demand. This metric summarizes the integrated outcome of atmospheric evaporative demand, water supply, vegetation regulation, and management during anomalous conditions, rather than isolating a single driver.

Demand-tracking ability varied across California's major land systems (Fig. 1a). In northern forests and many irrigated parts of the Central Valley, ETa generally rose as atmospheric evaporative demand increased. In drier interior landscapes dominated by grasslands and shrublands, ETa changed little or declined as demand increased. The representative grid cells show the monthly anomaly relationships underlying these spatial patterns. In the grassland and shrubland examples, ETa changed little and tended to decline as ETo increased, whereas ETa rose with ETo in the cropland and forest examples.

Land systems showed contrasting changes in demand-tracking ability with increasing drought severity (Fig. 1b). Forest and cropland medians remained positive even under severe drought, whereas grassland and shrubland medians declined from near zero under no-drought conditions to negative values under mild and severe drought. Pairwise comparisons further showed that mean demand-tracking ability was higher in forest and cropland than in grassland and shrubland within every drought class. Forest exceeded cropland only under no-drought conditions and did not differ detectably from cropland under mild or severe drought, while grassland and shrubland did not differ within any drought class (Supplementary Fig. S2). This separation is consistent with contrasting modes of water access. Grasslands and shrublands are more exposed to shallow or seasonal soil-water limitation, making weak or negative demand tracking consistent with water-limited conditions \cite{novickIncreasingImportanceAtmospheric2016, koningsSensitivityGrasslandProductivity2017}. Positive demand tracking in forests can be interpreted in terms of deeper rooting, stored-water access, and hydraulic diversity that can buffer ecosystem fluxes during dry periods \cite{andereggHydraulicDiversityForests2018}. In croplands, positive demand tracking points to managed water access and continued ETa on irrigated active fields. Together, these patterns show that drought did not uniformly suppress ETa responses across California, revealing where high atmospheric evaporative demand was translated into actual evapotranspiration and where water limitation weakened that response.

\begin{figure}[H]
\centering
\includegraphics[width=1.05\textwidth]{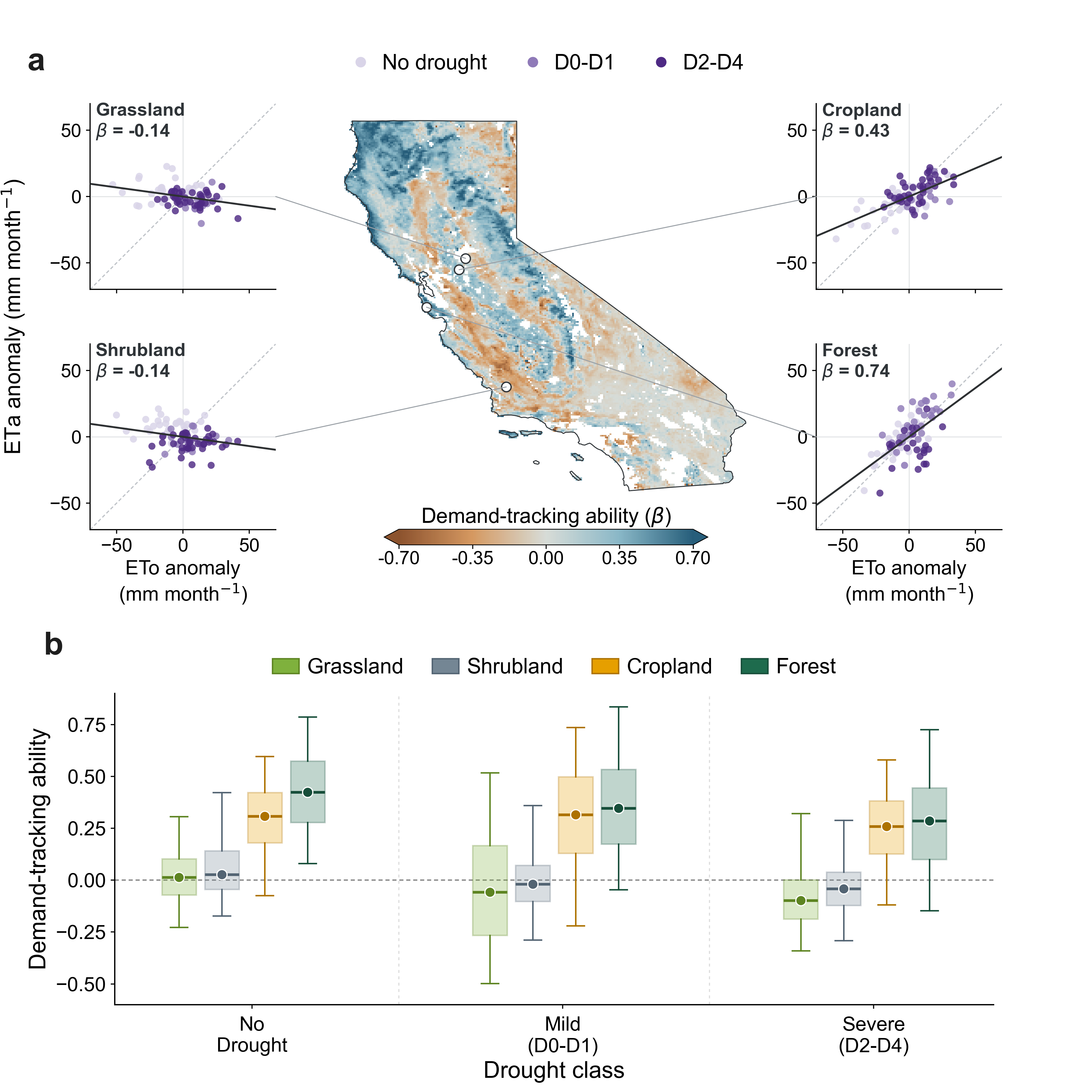}
\caption{\textbf{Demand tracking separates evapotranspiration responses across California land systems.} (a) Statewide demand-tracking ability from 2008 to 2024, calculated as the slope coefficient $\beta$ relating May–September ETa and ETo anomalies after subtracting each grid cell’s calendar-month mean. Insets show illustrative grassland, shrubland, cropland and forest grid cells indicated on the map. Points represent monthly anomalies coloured by US Drought Monitor class. Solid lines show fitted ETa–ETo relationships and dashed lines show the 1:1 relationship. (b) Demand-tracking ability across land systems and drought classes. Boxplots summarize all valid cell-level slope estimates passing predefined data-quality filters. Boxes show interquartile ranges, whiskers extend from the 5th to 95th percentiles and points mark medians.}
\end{figure}

\subsection*{Subsurface-storage context and crop-system accounting distinguish water-access modes}

Although forests and croplands both retained positive demand-tracking ability even under drought (Fig. 1), their similar \(ET_{\mathrm{a}}\) responses may not imply similar modes of water access. We examined water access in natural systems using three complementary comparisons among forests, grasslands, and shrublands. We focused this comparison on natural land systems because irrigation provides croplands with an additional water input. First, because precipitation can directly increase \(ET_{\mathrm{a}}\) within the same month, we re-estimated demand-tracking ability after accounting for same-month precipitation anomalies to test whether this immediate rainfall input could account for the stronger forest response (see Methods). We then used a previously published, snow-corrected estimate of apparent root-zone storage capacity, derived from a separate precipitation, evapotranspiration, and snow record, to characterize the relative capacity of the subsurface to store plant-accessible water through dry periods \citep{dralleTechnicalNoteAccounting2021}. Finally, we calculated mean May-September \(ET_{\mathrm{a}}-P\) to quantify how much growing-season \(ET_{\mathrm{a}}\) exceeded precipitation received during the same months. Positive values indicate that growing-season precipitation alone could not supply the observed \(ET_{\mathrm{a}}\), implying a contribution from water stored before the growing season or supplied laterally.

The forest contrast was consistent across all three comparisons. After precipitation adjustment, forest demand-tracking ability remained 0.28 higher than in grasslands and 0.19 higher than in shrublands (Fig. 2a,d). Apparent root-zone storage capacity was 166 mm and 113 mm greater, respectively (Fig. 2b,e), and growing-season \(ET_{\mathrm{a}}-P\) was 228 mm and 144 mm larger (Fig. 2c,f). All six 95\% confidence intervals excluded zero. The precipitation-adjusted contrast indicates that same-month precipitation variability did not account for the stronger forest response. The storage-capacity and \(ET_{\mathrm{a}}-P\) contrasts further show that forests had greater apparent capacity to store plant-accessible water and sustained more growing-season \(ET_{\mathrm{a}}\) than precipitation during those months alone could supply. Together, these contrasts show that positive demand-tracking ability over forests was consistent with access to subsurface storage.

\begin{figure}[H]
\centering
\includegraphics[width=1.05\textwidth]{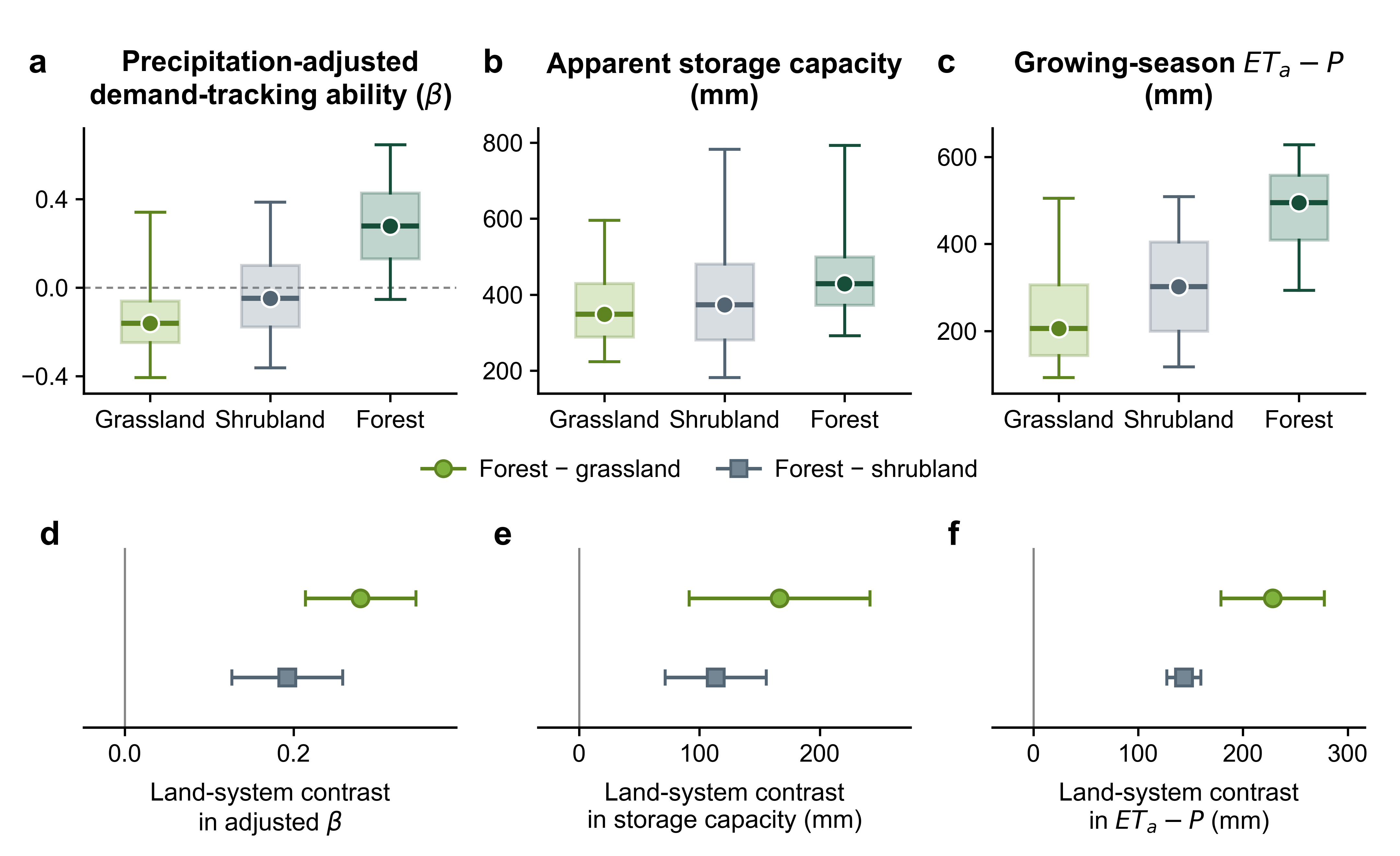}
\caption{\textbf{Forest responses are consistent with access to subsurface storage.} (a) Precipitation-adjusted demand-tracking ability, (b) snow-corrected apparent root-zone storage capacity, and (c) growing-season \(ET_{\mathrm{a}}-P\) across grassland, shrubland, and forest systems. Precipitation adjustment accounts for same-month precipitation anomalies. Boxplots summarize cell-level values. Boxes show interquartile ranges, whiskers extend from the 5th to 95th percentiles, and points mark medians. Forest–grassland and forest–shrubland contrasts in (d) precipitation-adjusted demand-tracking ability, (e) apparent root-zone storage capacity, and (f) growing-season \(ET_{\mathrm{a}}-P\). Points show estimated contrasts, and horizontal bars show 95\% confidence intervals. Positive values indicate higher values in forests.}
\end{figure}

We next used crop-system accounting to distinguish demand tracking over active cropland, crop-attributed fallow, and the whole crop system (Fig. 3). Within each 4-km grid cell, each crop system combined active cropland of the corresponding type with fallow area assigned according to the annual and perennial crop shares in the most recent preceding year with active cultivation. Demand tracking ability was then estimated for each whole crop system and separately for its active and fallow components. Median demand-tracking ability for annual and perennial crop systems remained positive across all drought classes, but the response did not decline monotonically with drought severity (Fig. 3a). From no drought to mild drought, the median remained nearly unchanged for annual systems and increased for perennial systems, before declining for both under severe drought. This pattern is consistent with managed water access buffering mild drought, followed by stronger water-supply constraints under severe drought. The crop-attributed fallow share of managed cropland increased from 10.5\% under no drought to 13.3\% under severe drought (Fig. 3b). Across D0–D4 observations, demand-tracking ability was substantially higher over active cropland than over crop-attributed fallow, with median values of 0.37 versus 0.06 for annual systems and 0.32 versus 0.08 for perennial systems (Fig. 3c).

Together, positive whole crop-system demand tracking (Fig. 3a), an increasing crop-attributed fallow share (Fig. 3b), and weaker demand tracking over crop-attributed fallow than over active cropland (Fig. 3c) show that active-field estimates alone do not fully characterize drought adjustment across the crop system. Figure 3d quantifies the resulting difference between active-field and whole crop-system estimates within matched grid cells. This difference was largest for annual systems under severe drought, whereas perennial point estimates were smaller across drought classes and approached zero under severe drought. Because crop-attributed fallow is inferred from yearly crop-type maps, it may include fields left fallow for crop rotations or other management reasons. Active-field and whole crop-system estimates can therefore differ even under no-drought conditions. During drought, positive demand tracking over active cropland is expected where irrigation supplies remain available because irrigation can sustain ETa as atmospheric evaporative demand rises. Yet this positive response on fields that remained active coincided with the rising crop-attributed fallow share. Thus, positive active-field ETa responses and a declining share of managed cropland under active cultivation represented opposing components of crop-system adjustment. Neither alone captured the full managed crop-system response.

\begin{figure}[H]
\centering
\includegraphics[width=1.05\textwidth]{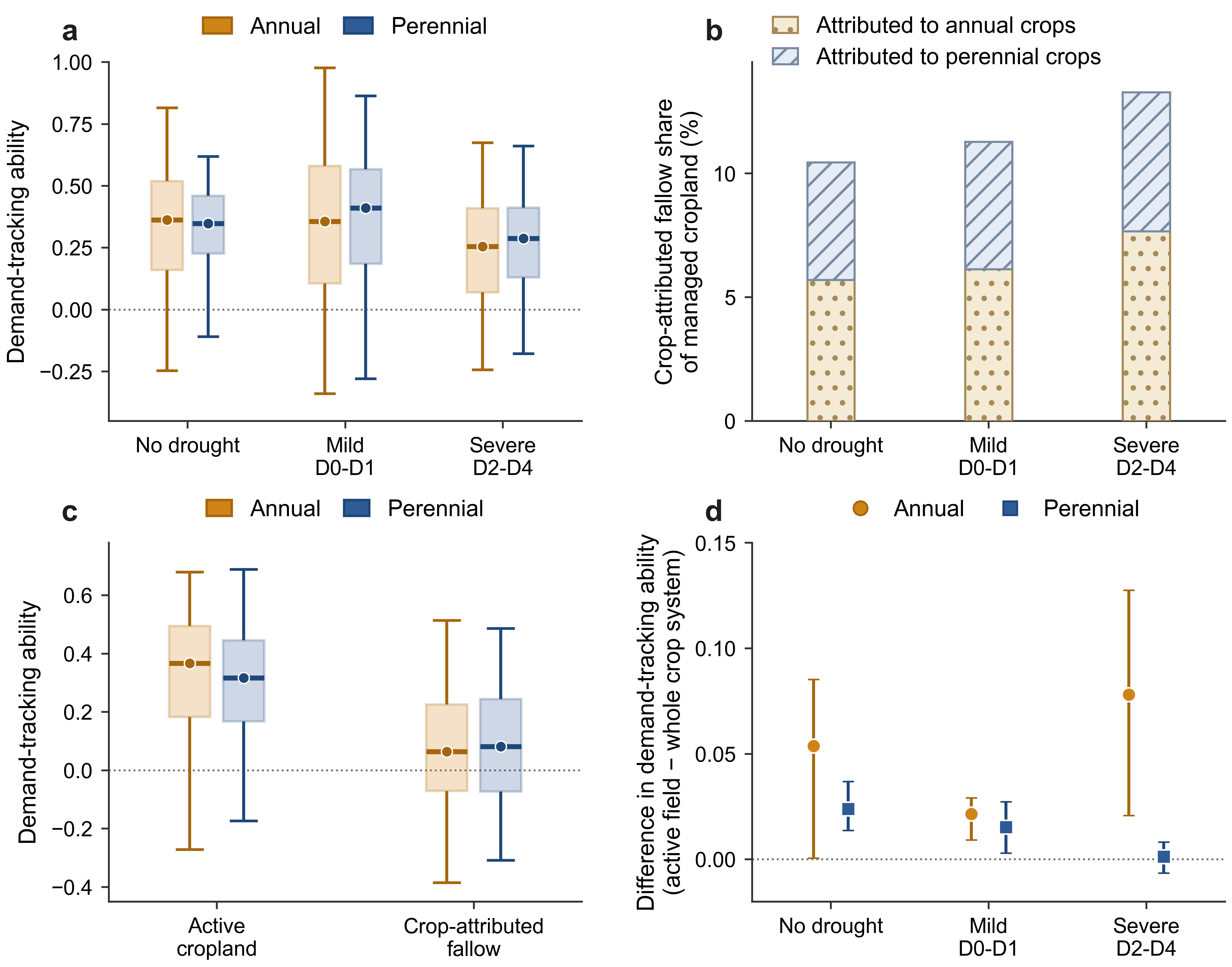}
\caption{\textbf{Crop-system accounting separates active-field and whole-system demand tracking.} (a) Whole crop-system demand-tracking ability for annual and perennial systems across drought classes. Annual and perennial crop systems comprise active cropland of the corresponding crop type plus crop-attributed fallow assigned to that system. Fallow was attributed in proportion to the annual and perennial crop shares observed in the most recent preceding year with active cultivation within each 4-km grid cell. Boxes show interquartile ranges, whiskers show 5th to 95th percentiles, and points mark medians. (b) Crop-attributed fallow as a share of managed cropland across drought classes, partitioned by attribution to annual and perennial crops. (c) Demand-tracking ability over active cropland and crop-attributed fallow for drought observations (D0-D4). Boxes show interquartile ranges of valid cell-level estimates, whiskers show 5th to 95th percentiles, and points mark medians. (d) Difference between active-field and whole crop-system demand-tracking ability for matched grid cells across drought classes. Points show mean paired differences, and error bars show 95\% confidence intervals accounting for spatial dependence. Positive values indicate stronger demand-tracking ability when only active fields are considered.}
\end{figure}

\subsection*{Demand tracking weakens with regional groundwater-storage loss}

Demand tracking also changed over time. We estimated demand-tracking ability separately for 2008-2015 and 2016-2024 to evaluate whether ETa-ETo coupling remained stable across repeated drought exposure (Fig. 4a,b). We quantified temporal change as demand-tracking ability in 2016-2024 minus that in 2008-2015, with negative values indicating late-period weakening and positive values indicating strengthening. Weakening was slightly more common than strengthening and was spatially uneven. Among 21,395 grid cells with valid estimates in both periods, the median change was $-0.02$, and 54.9\% showed weakening. Weakening was most apparent in parts of the Central Valley and along the Central and South Coast, whereas the North Coast remained stable or strengthened.

We then summarized this change by hydrologic region and compared it with the change in regional groundwater-storage anomaly (GWA), estimated by a component-residual storage approach adapted from Liu et al. \citep{liuGroundwaterDepletionCalifornias2022} (Methods; Fig. 4c,d). Regions with larger declines in GWA tended to show stronger declines in demand tracking. Across hydrologic regions, the association between mean GWA change and median demand-tracking change was positive because both quantities were defined as late minus early period (Pearson r = 0.71, p = 0.03). Tulare Lake had the most negative mean storage change, -133.6 mm, and also showed strong weakening in demand tracking.

At the hydrologic-region scale, this association links the weakening of demand tracking to broader groundwater storage-stress. GWA represents regional groundwater-storage change, whereas ETa, ETo and crop-system variables are analyzed on the 4-km resolution. The regional association is therefore interpreted as evidence that repeated drought and storage depletion coincide with weaker realized coupling between ETa and atmospheric evaporative demand, consistent with increasingly constrained water supplies. This interpretation aligns with previous evidence from the Central Valley that groundwater supports drought resilience in irrigated agriculture but has been depleted rapidly during the recent megadrought \cite{liuGroundwaterDepletionCalifornias2022}, and with global evidence that rapid groundwater-level declines are widespread in dry regions with extensive croplands \cite{jasechkoRapidGroundwaterDecline2024}.

\begin{figure}[H]
\centering
\includegraphics[width=1.05\textwidth]{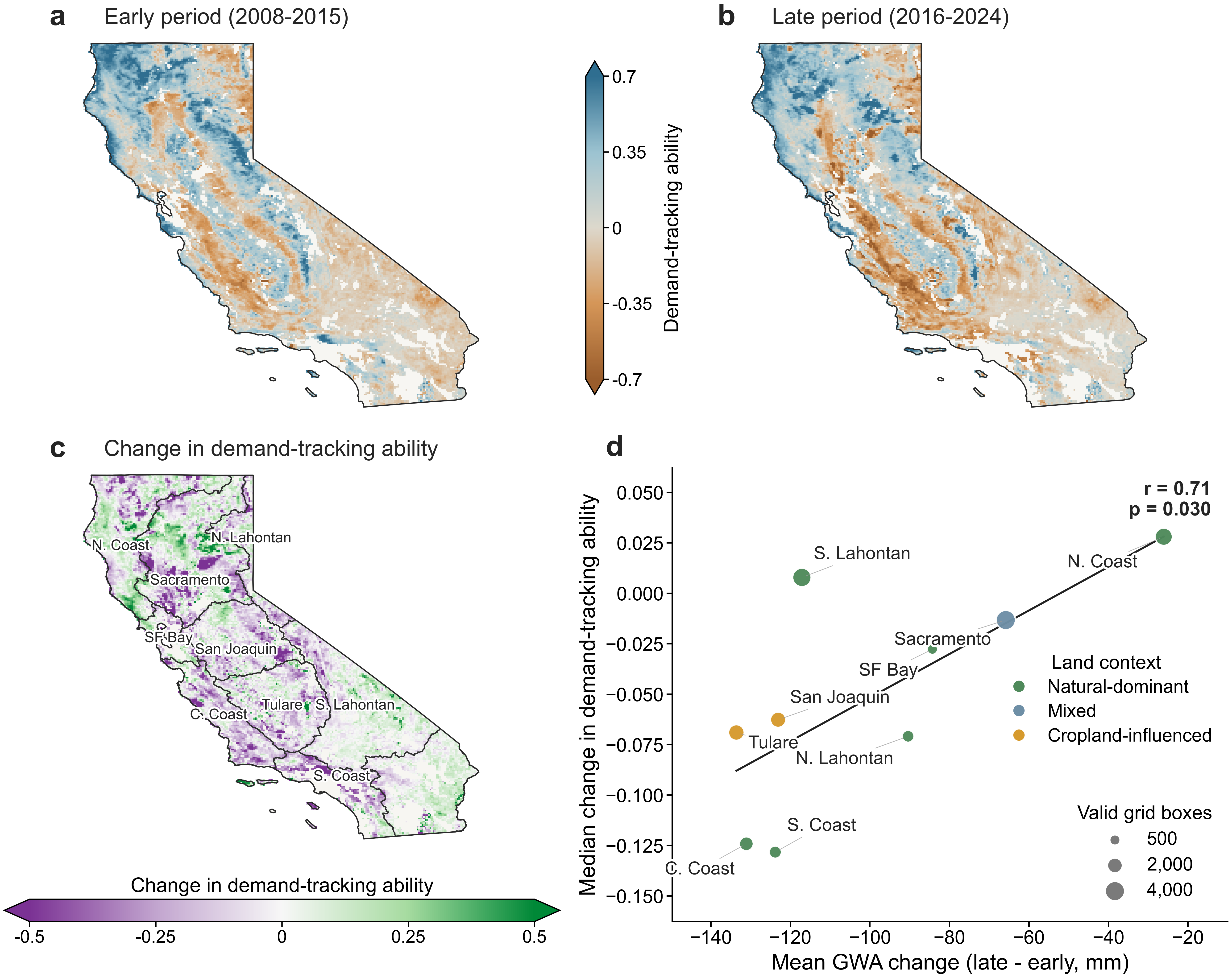}
\caption{\textbf{Demand tracking weakens unevenly and aligns with regional groundwater-storage loss.} (a,b) Demand-tracking ability estimated separately for 2008–2015 and 2016–2024. Both maps use the same colour scale and include grid cells with valid estimates in both periods. (c) Change in demand-tracking ability, calculated as late period minus early period. Negative values indicate late-period weakening and positive values indicate strengthening. (d) Hydrologic-region mean GWA change versus median change in demand-tracking ability across valid 4-km grid cells. Point colour indicates land-context class, and point size indicates the number of valid grid cells.}
\end{figure}

\subsection*{Severe-drought demand tracking separates natural and managed drought responses}

We next grouped severe-drought demand tracking into response states to summarize where water-limited decoupling, positive natural-system coupling, and managed crop-system adjustment occurred. Natural-system cells with positive severe-drought demand-tracking ability were classified as positive natural-system coupling, whereas those with zero or negative values were classified as water-limited decoupling. Cropland cells with zero or negative whole-system demand-tracking ability were classified as weak managed response. Among cells with positive whole-system demand-tracking ability, those with a large gap between active-cropland and whole-system tracking or a large increase in crop-attributed fallow share were classified as managed contraction. The remainder were classified as managed tracking. This response-state framework distinguishes ecosystem water limitation from managed agricultural responses under repeated drought stress.

The response-state map shows a clear spatial partition between natural and managed responses (Fig. 5a). Positive natural-system coupling was concentrated in northern and montane forested regions, while water-limited decoupling dominated many grassland, shrubland, and interior dryland landscapes. Managed states were concentrated in the Central Valley, where cropland responses reflected different combinations of maintained ETa coupling, fallowing, and stronger demand tracking over active cropland than across the whole crop system. This spatial structure is consistent with California's broader water-management context, in which irrigated agriculture can remain productive during drought through groundwater substitution and water allocation, but often with land idling and longer-term groundwater stress \citep{EconomicAnalysis2014, WaterFutureSan, liuGroundwaterDepletionCalifornias2022}.

Figure 5b summarizes the land- and crop-group composition within each response state. Grid cells classified as water-limited decoupling were predominantly grassland or shrubland. Managed tracking was dominated by perennial cropland, which accounted for 82\% of that state, while managed contraction contained a larger annual component, with annual systems contributing 57\%. Across hydrologic regions, northern coastal and montane regions were dominated by positive natural-system coupling, whereas southern, interior, and dryland regions were dominated by water-limited decoupling (Fig. 5c). The San Joaquin and Tulare regions contained the clearest mixture of managed tracking, managed contraction, and weak managed response, indicating that severe-drought cropland outcomes depend on both continued irrigation and cultivated-area adjustment rather than on ETa response alone. 

\begin{figure}[H]
\centering
\includegraphics[width=1.05\textwidth]{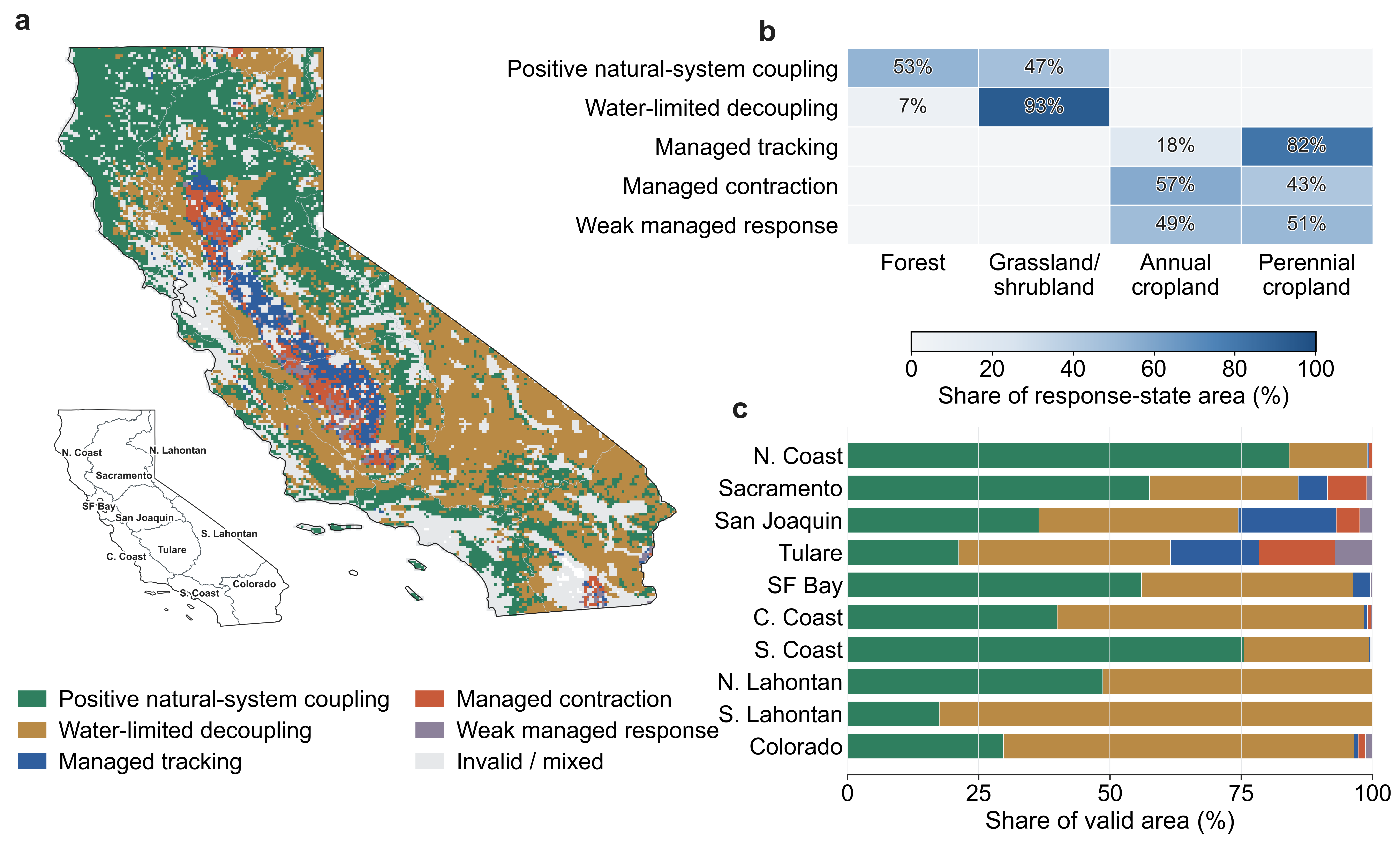}
\caption{\textbf{Severe-drought demand tracking defines diagnostic response states.} (a) Statewide response-state classification based on severe-drought demand-tracking ability. Each 4-km grid cell is assigned to the area-dominant valid land or crop-system component. Natural systems are classified as positive natural-system coupling or water-limited decoupling. Croplands are classified using whole crop-system severe-drought demand-tracking ability, the active-field versus whole crop-system tracking gap, and the increase in crop-attributed fallow share. Invalid or mixed cells are shown in light grey. (b) Land and crop-group composition within each response state. (c) Response-state composition within each hydrologic region, expressed as a share of valid classified area.}
\end{figure}

\section*{Discussion}

Rising atmospheric evaporative demand is increasingly shaping how drought is expressed on the land surface. In California, this demand-side pressure occurs within a heterogeneous landscape of forests, grasslands, shrublands and irrigated croplands, with long-term increases in growing-season VPD and ETo and repeated drought exposure during the 2008–2024 analysis window (Supplementary Fig. S1). Our results show that high ETo anomalies were not matched by a uniform ETa response. Forests and many irrigated croplands maintained positive demand tracking during drought, whereas grasslands and shrublands often shifted toward weak or negative tracking. Croplands also showed a managed response in which active fields remained coupled to atmospheric evaporative demand while the rest of the crop system adjusted through fallowing or crop transitions. Drought response is therefore expressed not only through water supply deficits, but through whether atmospheric evaporative demand, accessible water, vegetation function and management remain coupled during dry anomalies.

Previous work has established several parts of this problem. Atmospheric evaporative demand and soil moisture jointly regulate vegetation water use, surface conductance and carbon uptake, and rising VPD can intensify plant water stress during drought \citep{novickImpactsRisingVapour2024, novickIncreasingImportanceAtmospheric2016, grossiordPlantResponsesRising2020, humphreySoilMoistureAtmosphere2021}. Forest drought studies show that rooting depth, hydraulic traits, stored-water access, and changing resilience shape ecosystem responses to dry periods \citep{andereggHydraulicDiversityForests2018, fanHydrologicRegulationPlant2017, forzieriEmergingSignalsDeclining2022, mcdowellMechanismsWoodyplantMortality2022}. Agricultural studies have emphasized climate impacts on productivity, satellite ET monitoring, irrigation management and groundwater depletion in major food-producing regions \citep{ortiz-bobeaAnthropogenicClimateChange2021, volkAssessingAccuracyOpenET2024, liuGroundwaterDepletionCalifornias2022, jasechkoRapidGroundwaterDecline2024}. What remains unresolved is how water access, whether through subsurface storage in natural ecosystems or managed irrigation in croplands, can sustain ETa as atmospheric evaporative demand rises during drought. We address this question with a common anomaly-scale ETa–ETo metric and a cropland accounting boundary that distinguishes active-field persistence from whole crop-system adjustment.

Positive demand tracking therefore carries different hydrological meanings in forests and croplands. In forests, this response is consistent with deeper rooting, subsurface storage, and hydraulic diversity sustaining canopy fluxes during dry periods \citep{fanHydrologicRegulationPlant2017, andereggHydraulicDiversityForests2018, forzieriEmergingSignalsDeclining2022}. Positive forest demand tracking may therefore be duration-dependent: deep-root-zone storage provides a finite buffer, but prolonged drought can progressively deplete this reserve and reduce forests’ capacity to sustain ETa under high atmospheric evaporative demand \citep{gouldenCaliforniaForestDieOff2019}. In croplands, this response is instead consistent with managed water access and crop-system decisions, including crop selection, perennial investment, and which fields remain active during water stress \citep{liuGroundwaterDepletionCalifornias2022, WaterFutureSan, ortiz-bobeaAnthropogenicClimateChange2021}. A complementary spatial information partition provided broader context for these contrasting modes of water access. Among the static descriptors available here, land cover and hydroclimate made the largest contributions to natural-system patterns, whereas active/fallow structure and crop composition made the largest contributions to the cropland signal captured by the analysis (Supplementary Fig. S3). These distinct modes of water access also clarify how demand-tracking ability differs from established evaporative-stress indices. The Evaporative Stress Index (ESI) and related ratio-based metrics express ETa relative to potential or reference ET (PET or ETo), generally as standardized anomalies of that ratio \citep{andersonEvaluationDroughtIndices2011, yaoMonitoringDroughtConterminous2010, christianMethodologyFlashDrought2019}. Demand-tracking ability instead estimates how ETa itself varies with ETo anomalies. Positive demand tracking can therefore coexist with a declining ETa-to-ETo(or PET) ratio when ETa rises more slowly than ETo. The former captures an increase in ETa, whereas the latter indicates that ETa is meeting a smaller proportion of reference demand. Demand-tracking ability is thus a complementary response metric rather than a measure of drought severity. Complementary evaporation theory also cautions against interpreting this relationship causally because constrained ETa reduces evaporative cooling and can thereby increase potential or reference evaporation \citep{BouchetRJ1963a, brutsaertAdvectionaridityApproachEstimate1979}. Its hydrological meaning ultimately depends on whether the observed coupling is sustained by ecological water access or by management.

The managed component of this coupling becomes clearer when active-field and whole crop-system responses are considered together. If cropland ETa is evaluated only on active fields, positive tracking can be read as drought persistence. At the crop-system scale, however, the same positive tracking can coexist with fallowing and crop-area adjustment. Under severe drought, the active-field to whole crop-system difference was most pronounced for annual systems but approached zero for perennial systems, suggesting that annual systems adjusted more through fallowing, whereas the perennial response was expressed mainly within fields that remained active \citep{wongMultiscaleAssessmentAgricultural2021}. In active perennial fields, heat-related increases in crop water demand and irrigation constraints, including deficit irrigation, may both shape the observed ETa response, which could be further studied in the future \citep{moyersInvisibleWaterSurcharge2024,spinelliWaterStressCauses2016}. This distinction matters because satellite ET products are now accurate enough to support field-scale and regional water decisions \citep{fisherFutureEvapotranspirationGlobal2017, fisherEvapotranspirationEverywhereAll2026}, particularly in croplands, but ET estimates should still specify whether they refer to active fields alone or to the whole crop system before they can be interpreted as drought response or water-management signal \citep{volkAssessingAccuracyOpenET2024}. Complementing the regional groundwater-storage relationship in Fig. 4, croplands in areas with greater historical surface-water deliveries per irrigated acre showed stronger full-period demand tracking, while demand tracking weakened on average from the early to late period across all three delivery classes (Supplementary Fig. S4). Together, the groundwater-storage and surface-water-delivery patterns place cropland adjustment within a broader water-supply context, consistent with accelerated Central Valley groundwater depletion and global evidence of rapid groundwater declines in dry agricultural regions \citep{liuGroundwaterDepletionCalifornias2022, jasechkoRapidGroundwaterDecline2024,schipanskiMovingMeasurementGovernance2023}.

Several analyses support this interpretation while marking its limits. The broad statewide pattern was reproduced when demand-tracking ability was recalculated using individual OpenET component models, although magnitudes and local spatial details varied (Supplementary Fig. S5). Most component estimates showed moderate to strong agreement with the ensemble, with weaker agreement for SIMS (Supplementary Fig. S6). These comparisons are interpreted as internal product-sensitivity checks rather than as independent validation of ETa. The response-state map should be read as a synthesis of predefined diagnostic rules rather than as a process model. Source-derived fallow is also an accounting state rather than a homogeneous land-cover class, and it can include rotational fallow, persistent low-activity land, and drought-related idling. Future applications would benefit from combining demand tracking with annual surface-water deliveries, groundwater pumping, irrigation district allocations, crop economics, and local groundwater observations, especially where the goal is to separate active-field ET responses from whole crop-system adjustment.

Demand tracking adds a diagnostic layer to satellite-based water accounting. Rather than treating ET only as a consumptive-use total, it asks whether atmospheric evaporative demand is realized as ETa, constrained by water limitation or redistributed through management. Such applications are becoming more feasible as satellite ET products improve and as water management increasingly requires spatially explicit accounting of consumptive use \citep{volkAssessingAccuracyOpenET2024, meltonOpenETFillingCritical2022, knipperComparativeAnalysisOpenET2024, wongMultiscaleAssessmentAgricultural2021}. They are also increasingly relevant as hydroclimatic variability intensifies and groundwater depletion remains concentrated in many dry agricultural regions \citep{swainIncreasingPrecipitationVolatility2018, jasechkoRapidGroundwaterDecline2024}. Demand tracking does not replace process-based hydrology, irrigation accounting or groundwater monitoring. Instead, applied alongside land-system and water-supply context, it provides a common way to compare where ETa–demand coupling is sustained, where it weakens under water limitation and where managed systems maintain ETa on active fields while fallow expands. More broadly, this perspective can extend beyond California to other water-limited regions where natural ecosystems and managed agriculture face rising evaporative demand, provided that ET, atmospheric evaporative demand, and land-system context can be evaluated consistently.

\section*{Methods}

\subsection*{Data sources}

\textbf{Evapotranspiration, atmospheric evaporative demand, and drought} Actual evapotranspiration (ETa) was obtained from the OpenET ensemble \citep{volkAssessingAccuracyOpenET2024}. Reference evapotranspiration (ETo), used as the standardized atmospheric evaporative-demand variable, was taken from gridMET \citep{Abatzoglou2013}. GridMET vapour pressure deficit, precipitation and temperature were also used as hydroclimatic covariates. Drought conditions were taken from the U.S. Drought Monitor and grouped into no drought, mild drought (D0-D1) and severe drought (D2-D4) \citep{SvobodaEtAl2002}. Severe-drought months were used for response-state classification.

\textbf{Land cover, crop systems, and surface variables} Land cover was derived from the National Land Cover Database (NLCD) 2019 product \citep{NationalLandCover2023,YangEtAl2018}. Within each 4-km grid cell, we defined the dominant land-cover class as the NLCD class occupying the largest share of valid NLCD pixels, provided that this share was at least 50\%. Cells not meeting this threshold were classified as mosaics. Natural-system analyses retained cells classified as forest, grassland or shrubland. Crop type, annual and perennial crop groups, and fallow states were derived from the annual USDA Cropland Data Layer and associated crop classifications \citep{BoryanEtAl2011}. Soil available water capacity was derived from gridded soil-survey data, and elevation and slope were derived from national elevation data \citep{SoilSurveyStaffGSSURGO,3DElevationProgram}. Categorical datasets were aggregated to the 4-km grid as area fractions, and continuous datasets were averaged or resampled to the same grid before analysis.

\textbf{Apparent root-zone storage capacity} Snow-corrected apparent root-zone storage capacity was obtained from a nominally 1-km dataset covering the continental United States \citep{dralleTechnicalNoteAccounting2021}. This product extends the cumulative root-zone storage-deficit approach based on precipitation and evapotranspiration fluxes \citep{wang-erlandssonGlobalRootZone2016} by accounting for snow cover. We used Band 2, \(S_{R,\mathrm{snow\,corr}}\), reported in millimetres and derived by continuously tracking root-zone storage deficits from water year 2003 through water year 2017. The raster was aggregated to the 4-km analysis grid using the exact spherical area of each source-target overlap as the weight. Non-finite source pixels were excluded from the weighted mean, whereas finite zero values were retained. Valid coverage was defined as the finite Band-2 overlap area divided by the full target-cell area, and cells were retained when this fraction was at least 50\%. The resulting quantity is intended as a flux-derived lower-bound estimate of the root-zone storage capacity accessed by vegetation over the source record, rather than a direct measurement of time-varying subsurface water storage or an attribution of groundwater uptake or rooting depth.

\subsection*{Demand-tracking analysis}

Demand-tracking ability was defined from deseasonalized growing-season (May-September) anomalies of ETa and ETo. For each 4-km grid cell \(i\), month \(t\) and growing-season calendar month \(m\), anomalies were calculated as

\begin{equation}
\Delta X_{i,t} = X_{i,t} - \overline{X}_{i,m},
\qquad
X \in \{\mathrm{ETa}, \mathrm{ETo}\},
\end{equation}

\noindent where overbars denote the 2008-2024 mean for the same grid cell and calendar month. This removes the local seasonal cycle and focuses the analysis on departures from expected May-September conditions.

For each analysis unit, demand-tracking ability was estimated as the slope between ETa and ETo anomalies:

\begin{equation}
\Delta \mathrm{ETa}_{i,t}
=
\alpha_i
+
\beta_i \Delta \mathrm{ETo}_{i,t}
+
\epsilon_{i,t}.
\end{equation}

\noindent The coefficient \(\beta_i\) is the demand-tracking ability. Positive values indicate that ETa anomalies tended to increase during high-ETo anomalies, whereas values near or below zero indicate weak or reversed ETa-demand coupling. Because \(\beta_i\) is estimated from observed anomalies, it is interpreted as an empirical coupling metric rather than as an experimental estimate of the causal effect of ETo on ETa. The intercept \(\alpha_i\) was estimated freely and was not constrained to zero.

Unless otherwise noted, a slope estimate was retained when \(\beta\) was finite, at least ten monthly observations were available, at least two years contributed observations, the standard deviation of \(\Delta \mathrm{ETo}\) was at least 5 mm/month, and the regression succeeded. Drought-class slopes were estimated using the same anomaly definition after grouping monthly observations by U.S. Drought Monitor class. Land-system estimates in Fig. 1 used the dominant 4-km land-cover class to assign grid cells to cropland, forest, grassland or shrubland.

For the natural-system comparison in Fig. 2, we additionally estimated precipitation-adjusted demand-tracking ability as

\begin{equation}
\Delta ET_{\mathrm{a},i,t}
=
\alpha_i
+
\beta_{i,\mathrm{ETo}\mid P}\Delta ETo_{i,t}
+
\gamma_i\Delta P_{i,t}
+
\epsilon_{i,t}.
\end{equation}

The coefficient \(\beta_{i,\mathrm{ETo}\mid P}\) represents the relationship between \(ET_{\mathrm{a}}\) and \(ETo\) anomalies after accounting for same-month precipitation anomalies. The model was estimated from all May-September observations without drought-class or period stratification. Estimates were retained when at least ten observations and two years were available, the standard deviations of \(\Delta ETo\) and \(\Delta P\) were at least 5 and 1 mm month\(^{-1}\), respectively, the variance inflation factor did not exceed 10, and all model coefficients were finite.

For the temporal comparison, we repeated the same calculation after splitting the record into an early period, 2008-2015, and a late period, 2016-2024. Period-specific ETa and ETo anomalies were calculated within each grid cell and calendar month using the corresponding period mean. Temporal change was defined as
\(\Delta \beta_i = \beta^{\mathrm{late}}_i - \beta^{\mathrm{early}}_i\), with negative values indicating weaker ETa-demand coupling in the late period.

\subsection*{Contrasts among natural land systems}

We compared forests with grasslands and shrublands for three outcomes: precipitation-adjusted demand-tracking ability, snow-corrected apparent root-zone storage capacity, and mean May-September \(ET_{\mathrm a}-P\). All outcomes were analysed using the same set of 4-km grid cells with valid values for all three outcomes.

For each outcome \(Y_i\), we fitted an unweighted ordinary least-squares model at the grid-cell level:

\begin{equation}
Y_i=\mu+\gamma_{\mathrm S}\mathbb{I}(L_i=\mathrm{Shrubland})+\gamma_{\mathrm F}\mathbb{I}(L_i=\mathrm{Forest})+\lambda_{b(i)}+\varepsilon_i .
\end{equation}

Here, \(L_i\) denotes the dominant natural land system, and \(b(i)\) denotes the predefined \(1^\circ \times 1^\circ\) latitude-longitude spatial block containing cell \(i\). These blocks were approximately 100 km across. Grassland was the reference class. The block-specific intercepts \(\lambda_{b(i)}\) allowed mean outcome levels to vary among blocks, thereby estimating land-system contrasts within broad regional settings. The coefficient \(\gamma_{\mathrm F}\) represents the Forest--Grassland contrast, whereas \(\gamma_{\mathrm F}-\gamma_{\mathrm S}\) represents the Forest--Shrubland contrast.

Standard errors were clustered by the same spatial blocks to account for spatial dependence among grid cells. Finite-sample and degrees-of-freedom corrections were applied, and inference used a \(t\) distribution. We report pointwise two-sided 95\% confidence intervals. The six outcome-by-contrast \(P\) values were adjusted using the Holm method.

\subsection*{Crop-system accounting}
Cropland demand-tracking ability was estimated separately for annual and perennial crop systems under active-field and whole crop-system accounting. Within each grid cell, crop-attributed fallow was assigned to annual and perennial systems in proportion to their shares in the most recent preceding year with active cultivation. Fallow with no identifiable preceding active-crop source was retained as persistent or unclear fallow and was not assigned to either system.

For each crop system and drought class, demand-tracking ability was estimated using equation (2), with active-crop and crop-attributed-fallow observations weighted by their respective area fractions. Active-field estimates used only active-crop observations, whereas whole crop-system estimates combined active-crop and corresponding crop-attributed-fallow observations in a single area-weighted regression.

For Fig. 3d, active-field and whole crop-system estimates were matched by grid cell, crop system, and drought class, retaining only pairs for which both estimates passed the predefined quality-control criteria. For each matched cell, we calculated \(\beta^{\mathrm{active}}_{i,c,d} -\beta^{\mathrm{whole}}_{i,c,d}\), where \(i\), \(c\), and \(d\) denote grid cell, crop system, and drought class, respectively. Figure 3d reports the mean of these paired cell-level differences. To account for spatial dependence, 95\% confidence intervals were estimated from 1,000 percentile bootstrap replicates in which approximately 100-km spatial blocks were resampled with replacement and all matched cells within each sampled block were retained. The 2.5th and 97.5th percentiles of the bootstrap distribution defined the interval bounds.

\subsection*{Groundwater-storage context}

Regional groundwater-storage anomaly (GWA) was estimated using a component-residual storage approach adapted from Liu et al. \citep{liuGroundwaterDepletionCalifornias2022}. Total water storage anomaly (TWSA) was taken from the JPL GRACE/GRACE-FO mascon product \citep{JPLGRACEMascon}. Soil moisture anomaly (SMA) was calculated from North American Land Data Assimilation System phase 2 (NLDAS-2) soil moisture ensemble \citep{xiaContinentalscaleWaterEnergy2012}, snow water equivalent anomaly (SWEA) from SNODAS end-of-month snow water equivalent \citep{NOHRSCSNODAS}, and surface-water anomaly (SWA) from California Data Exchange Center monthly reservoir storage \citep{CDEC}. All storage components were converted to equivalent water-height anomalies before hydrologic-region aggregation.

For hydrologic region \(r\) and month \(t\), GWA was calculated as

\begin{equation}
\mathrm{GWA}_{r,t}
=
\mathrm{TWSA}_{r,t}
-
\mathrm{SMA}_{r,t}
-
\mathrm{SWEA}_{r,t}
-
\mathrm{SWA}_{r,t},
\end{equation}

\noindent where TWSA, SMA, SWEA and SWA are total water storage, soil moisture, snow water equivalent and surface-water storage anomalies, respectively. GWA was used only as a regional storage-context variable because GRACE-based storage anomalies represent broad mass changes, whereas ETa, ETo, land-cover and crop-system variables were analysed on the 4-km grid.

For the temporal comparison, regional groundwater-storage change was calculated as late-period mean GWA minus early-period mean GWA. The association between regional GWA change and median demand-tracking change was evaluated using Pearson correlation.

\subsection*{Severe-drought response states}

Response states were classified using severe-drought (D2-D4) demand-tracking estimates. Candidate components were forest, grassland, and shrubland for natural systems, and annual and perennial crop systems for croplands. Each 4-km grid cell was assigned to the valid candidate with the largest area fraction. Cells assigned to natural systems were classified as positive natural-system coupling when severe-drought demand-tracking ability was positive and as water-limited decoupling when it was zero or negative.

Cropland response states were classified using three quantities: whole-system severe-drought demand-tracking ability, the active-field versus whole crop-system demand-tracking ability gap ($\beta_{\mathrm{active}}-\beta_{\mathrm{whole}}$), and the increase in crop-attributed fallow share. The classification workflow is summarized in Supplementary Fig. S7. The increase in crop-attributed fallow share was calculated as its mean value during D2-D4 minus its mean under no-drought conditions for the same crop system. For each of the two adjustment metrics, we calculated an area-weighted 67th percentile across valid cropland components, so that the upper third of cropland area defined high adjustment. A separate threshold was calculated for each metric. Cropland cells with non-positive whole-system demand-tracking ability were classified as weak managed response. Among cells with positive whole-system demand-tracking ability, cells were classified as managed tracking when neither adjustment metric was high and as managed contraction when either metric was high.

\backmatter

\section*{Data availability} All input datasets used in this study are publicly available from the following official data portals and repositories: OpenET (ensemble and component-model evapotranspiration), \url{https://developers.google.com/earth-engine/datasets/publisher/openet}; gridMET (reference evapotranspiration and meteorological variables), \url{https://www.climatologylab.org/gridmet.html}; U.S. Drought Monitor (drought classifications), \url{https://droughtmonitor.unl.edu/DmData/GISData.aspx}; National Land Cover Database (land cover), \url{https://www.mrlc.gov/data}; USDA Cropland Data Layer (annual crop classifications), \url{https://www.nass.usda.gov/Research_and_Science/Cropland/Release/index.php}; USDA-NRCS gSSURGO (soil properties), \url{https://www.nrcs.usda.gov/resources/data-and-reports/gridded-soil-survey-geographic-gssurgo-database}; USGS 3D Elevation Program (elevation), \url{https://www.usgs.gov/the-national-map-data-delivery/gis-data-download}; HydroShare (snow-corrected apparent root-zone storage capacity), \url{https://www.hydroshare.org/resource/ee45c2f5f13042ca85bcb86bbfc9dd37/}; JPL GRACE/GRACE-FO mascon data (terrestrial water-storage anomalies), \url{https://podaac.jpl.nasa.gov/dataset/TELLUS_GRAC-GRFO_MASCON_CRI_GRID_RL06.3_V4}; NLDAS-2 (soil moisture), \url{https://ldas.gsfc.nasa.gov/nldas/v2/models}; SNODAS Version 1 (snow water equivalent), \url{https://nsidc.org/data/g02158/versions/1}; California Data Exchange Center (reservoir storage), \url{https://cdec.water.ca.gov/}; and Public Policy Institute of California surface-water availability datasets for the San Joaquin Valley, \url{https://www.ppic.org/data-set/ppic-san-joaquin-valley-surface-water-availability/}, and the Sacramento Valley and Delta, \url{https://www.ppic.org/data-set/ppic-sacramento-valley-and-delta-surface-water-availability/}. Processed grid-cell-level datasets, regional storage time series, and analysis-ready files generated during this study are available from the corresponding author upon reasonable request.

\bmhead{Acknowledgements}

YZ acknowledges support from the UC Irvine Climate Collaboration. JBF was supported in part by NASA ECOSTRESS Science and Applications Team (ESAT) (80NSSC23K0309), OpenET, and NSF Division of Earth Sciences (2012893) through CUAHSI and the USGS John Wesley Powell Center for Analysis and Synthesis. AEB was supported in part by the National Oceanic and Atmospheric Administration's National Integrated Drought Information System.

\bibliography{references}

\begin{thebibliography}{10}
\expandafter\ifx\csname url\endcsname\relax
  \def\url#1{\burl{#1}}\fi
\expandafter\ifx\csname urlprefix\endcsname\relax\def\urlprefix{URL }\fi
\providecommand{\bibinfo}[2]{#2}
\providecommand{\eprint}[2][]{\url{#2}}
\providecommand{\doi}[1]{\url{https://doi.org/#1}}
\bibcommenthead

\bibitem{bergLandAtmosphereFeedbacks2016}
\bibinfo{author}{Berg, A.} \emph{et~al.}
\newblock \bibinfo{title}{Land–atmosphere feedbacks amplify aridity increase over land under global warming}.
\newblock \emph{\bibinfo{journal}{Nature Climate Change}} \textbf{\bibinfo{volume}{6}}, \bibinfo{pages}{869--874} (\bibinfo{year}{2016}).

\bibitem{denissenWidespreadShiftEcosystem2022}
\bibinfo{author}{Denissen, J. M.~C.} \emph{et~al.}
\newblock \bibinfo{title}{Widespread shift from ecosystem energy to water limitation with climate change}.
\newblock \emph{\bibinfo{journal}{Nature Climate Change}} \textbf{\bibinfo{volume}{12}}, \bibinfo{pages}{677--684} (\bibinfo{year}{2022}).

\bibitem{gebrechorkosWarmingAcceleratesGlobal2025}
\bibinfo{author}{Gebrechorkos, S.~H.} \emph{et~al.}
\newblock \bibinfo{title}{Warming accelerates global drought severity}.
\newblock \emph{\bibinfo{journal}{Nature}} \textbf{\bibinfo{volume}{642}}, \bibinfo{pages}{628--635} (\bibinfo{year}{2025}).

\bibitem{novickImpactsRisingVapour2024}
\bibinfo{author}{Novick, K.~A.} \emph{et~al.}
\newblock \bibinfo{title}{The impacts of rising vapour pressure deficit in natural and managed ecosystems}.
\newblock \emph{\bibinfo{journal}{Plant, Cell \& Environment}} \textbf{\bibinfo{volume}{47}}, \bibinfo{pages}{3561--3589} (\bibinfo{year}{2024}).

\bibitem{seneviratneInvestigatingSoilMoisture2010}
\bibinfo{author}{Seneviratne, S.~I.} \emph{et~al.}
\newblock \bibinfo{title}{Investigating soil moisture–climate interactions in a changing climate: a review}.
\newblock \emph{\bibinfo{journal}{Earth-Science Reviews}} \textbf{\bibinfo{volume}{99}}, \bibinfo{pages}{125--161} (\bibinfo{year}{2010}).

\bibitem{budykoClimateLife1974}
\bibinfo{author}{Budyko, M.~I.}
\newblock \emph{\bibinfo{title}{Climate and life}} No.~\bibinfo{number}{18} in \bibinfo{series}{International geophysics series} (\bibinfo{publisher}{Academic Press}, \bibinfo{address}{New York}, \bibinfo{year}{1974}).

\bibitem{novickIncreasingImportanceAtmospheric2016}
\bibinfo{author}{Novick, K.~A.} \emph{et~al.}
\newblock \bibinfo{title}{The increasing importance of atmospheric demand for ecosystem water and carbon fluxes}.
\newblock \emph{\bibinfo{journal}{Nature Climate Change}} \textbf{\bibinfo{volume}{6}}, \bibinfo{pages}{1023--1027} (\bibinfo{year}{2016}).

\bibitem{jungRecentDeclineGlobal2010}
\bibinfo{author}{Jung, M.} \emph{et~al.}
\newblock \bibinfo{title}{Recent decline in the global land evapotranspiration trend due to limited moisture supply}.
\newblock \emph{\bibinfo{journal}{Nature}} \textbf{\bibinfo{volume}{467}}, \bibinfo{pages}{951--954} (\bibinfo{year}{2010}).

\bibitem{allenrichardg.CropEvapotranspirationGuidelines1998}
\bibinfo{author}{{Allen, Richard G.}}, \bibinfo{author}{{Pereira, Luis S.}}, \bibinfo{author}{{Raes, Dirk}} \& \bibinfo{author}{{Smith, Martin}}.
\newblock \bibinfo{title}{Crop evapotranspiration: {Guidelines} for computing crop water requirements}.
\newblock \bibinfo{type}{{FAO} {Irrigation} and {Drainage} {Paper}} \bibinfo{number}{56}, \bibinfo{institution}{Food and Agriculture Organization of the United Nations}, \bibinfo{address}{Rome} (\bibinfo{year}{1998}).

\bibitem{fisherComeHomePotential2011}
\bibinfo{author}{Fisher, J.~B.}, \bibinfo{author}{Whittaker, R.~J.} \& \bibinfo{author}{Malhi, Y.}
\newblock \bibinfo{title}{{ET} come home: potential evapotranspiration in geographical ecology}.
\newblock \emph{\bibinfo{journal}{Global Ecology and Biogeography}} \textbf{\bibinfo{volume}{20}}, \bibinfo{pages}{1--18} (\bibinfo{year}{2011}).

\bibitem{grossiordPlantResponsesRising2020}
\bibinfo{author}{Grossiord, C.} \emph{et~al.}
\newblock \bibinfo{title}{Plant responses to rising vapor pressure deficit}.
\newblock \emph{\bibinfo{journal}{New Phytologist}} \textbf{\bibinfo{volume}{226}}, \bibinfo{pages}{1550--1566} (\bibinfo{year}{2020}).

\bibitem{BouchetRJ1963a}
\bibinfo{author}{Bouchet, R.}
\newblock \bibinfo{title}{ in \textit{Evapotranspiration réelle et potentielle, signification climatique}} \emph{\bibinfo{booktitle}{General {Assembly} of {Berkeley}, {Transactions}, {Vol}. 2: {Evaporation}}} No.~\bibinfo{number}{62} in \bibinfo{series}{{IAHS} {Publication}}, \bibinfo{pages}{134 -- 142} (\bibinfo{publisher}{International Association of Scientific Hydrology}, \bibinfo{address}{Berkeley, CA}, \bibinfo{year}{1963}).

\bibitem{brutsaertAdvectionaridityApproachEstimate1979}
\bibinfo{author}{Brutsaert, W.} \& \bibinfo{author}{Stricker, H.}
\newblock \bibinfo{title}{An advection-aridity approach to estimate actual regional evapotranspiration}.
\newblock \emph{\bibinfo{journal}{Water Resources Research}} \textbf{\bibinfo{volume}{15}}, \bibinfo{pages}{443--450} (\bibinfo{year}{1979}).

\bibitem{fisherGlobalEstimatesLand2008}
\bibinfo{author}{Fisher, J.~B.}, \bibinfo{author}{Tu, K.~P.} \& \bibinfo{author}{Baldocchi, D.~D.}
\newblock \bibinfo{title}{Global estimates of the land–atmosphere water flux based on monthly {AVHRR} and {ISLSCP}-{II} data, validated at 16 {FLUXNET} sites}.
\newblock \emph{\bibinfo{journal}{Remote Sensing of Environment}} \textbf{\bibinfo{volume}{112}}, \bibinfo{pages}{901--919} (\bibinfo{year}{2008}).

\bibitem{mcdowellDarcysLawPredicts2015}
\bibinfo{author}{McDowell, N.~G.} \& \bibinfo{author}{Allen, C.~D.}
\newblock \bibinfo{title}{Darcy's law predicts widespread forest mortality under climate warming}.
\newblock \emph{\bibinfo{journal}{Nature Climate Change}} \textbf{\bibinfo{volume}{5}}, \bibinfo{pages}{669--672} (\bibinfo{year}{2015}).

\bibitem{andereggHydraulicDiversityForests2018}
\bibinfo{author}{Anderegg, W. R.~L.} \emph{et~al.}
\newblock \bibinfo{title}{Hydraulic diversity of forests regulates ecosystem resilience during drought}.
\newblock \emph{\bibinfo{journal}{Nature}} \textbf{\bibinfo{volume}{561}}, \bibinfo{pages}{538--541} (\bibinfo{year}{2018}).

\bibitem{mccormickWidespreadWoodyPlant2021}
\bibinfo{author}{McCormick, E.~L.} \emph{et~al.}
\newblock \bibinfo{title}{Widespread woody plant use of water stored in bedrock}.
\newblock \emph{\bibinfo{journal}{Nature}} \textbf{\bibinfo{volume}{597}}, \bibinfo{pages}{225--229} (\bibinfo{year}{2021}).

\bibitem{volkAssessingAccuracyOpenET2024}
\bibinfo{author}{Volk, J.~M.} \emph{et~al.}
\newblock \bibinfo{title}{Assessing the accuracy of {OpenET} satellite-based evapotranspiration data to support water resource and land management applications}.
\newblock \emph{\bibinfo{journal}{Nature Water}} \textbf{\bibinfo{volume}{2}}, \bibinfo{pages}{193--205} (\bibinfo{year}{2024}).

\bibitem{boserFieldscaleCropWater2024}
\bibinfo{author}{Boser, A.} \emph{et~al.}
\newblock \bibinfo{title}{Field-scale crop water consumption estimates reveal potential water savings in {California} agriculture}.
\newblock \emph{\bibinfo{journal}{Nature Communications}} \textbf{\bibinfo{volume}{15}}, \bibinfo{pages}{2366} (\bibinfo{year}{2024}).

\bibitem{Abatzoglou2013}
\bibinfo{author}{Abatzoglou, J.~T.}
\newblock \bibinfo{title}{Development of gridded surface meteorological data for ecological applications and modelling}.
\newblock \emph{\bibinfo{journal}{International Journal of Climatology}} \textbf{\bibinfo{volume}{33}}, \bibinfo{pages}{121--131} (\bibinfo{year}{2013}).

\bibitem{SvobodaEtAl2002}
\bibinfo{author}{Svoboda, M.} \emph{et~al.}
\newblock \bibinfo{title}{The {Drought} {Monitor}}  (\bibinfo{year}{2002}).

\bibitem{williamsRapidIntensificationEmerging2022}
\bibinfo{author}{Williams, A.~P.}, \bibinfo{author}{Cook, B.~I.} \& \bibinfo{author}{Smerdon, J.~E.}
\newblock \bibinfo{title}{Rapid intensification of the emerging southwestern {North} {American} megadrought in 2020–2021}.
\newblock \emph{\bibinfo{journal}{Nature Climate Change}} \textbf{\bibinfo{volume}{12}}, \bibinfo{pages}{232--234} (\bibinfo{year}{2022}).

\bibitem{marstonDroughtImpactsWater2017}
\bibinfo{author}{Marston, L.} \& \bibinfo{author}{Konar, M.}
\newblock \bibinfo{title}{Drought impacts to water footprints and virtual water transfers of the {Central} {Valley} of {California}}.
\newblock \emph{\bibinfo{journal}{Water Resources Research}} \textbf{\bibinfo{volume}{53}}, \bibinfo{pages}{5756--5773} (\bibinfo{year}{2017}).

\bibitem{yangImpactIrrigationCalifornia2017}
\bibinfo{author}{Yang, Z.} \emph{et~al.}
\newblock \bibinfo{title}{Impact of {Irrigation} over the {California} {Central} {Valley} on {Regional} {Climate}}  (\bibinfo{year}{2017}).

\bibitem{mcdermidIrrigationEarthSystem2023}
\bibinfo{author}{McDermid, S.} \emph{et~al.}
\newblock \bibinfo{title}{Irrigation in the {Earth} system}.
\newblock \emph{\bibinfo{journal}{Nature Reviews Earth \& Environment}} \textbf{\bibinfo{volume}{4}}, \bibinfo{pages}{435--453} (\bibinfo{year}{2023}).

\bibitem{liuGroundwaterDepletionCalifornias2022}
\bibinfo{author}{Liu, P.-W.} \emph{et~al.}
\newblock \bibinfo{title}{Groundwater depletion in {California}’s {Central} {Valley} accelerates during megadrought}.
\newblock \emph{\bibinfo{journal}{Nature Communications}} \textbf{\bibinfo{volume}{13}}, \bibinfo{pages}{7825} (\bibinfo{year}{2022}).

\bibitem{jasechkoRapidGroundwaterDecline2024}
\bibinfo{author}{Jasechko, S.} \emph{et~al.}
\newblock \bibinfo{title}{Rapid groundwater decline and some cases of recovery in aquifers globally}.
\newblock \emph{\bibinfo{journal}{Nature}} \textbf{\bibinfo{volume}{625}}, \bibinfo{pages}{715--721} (\bibinfo{year}{2024}).

\bibitem{mienoAquiferDepletionExacerbates2024}
\bibinfo{author}{Mieno, T.}, \bibinfo{author}{Foster, T.}, \bibinfo{author}{Kakimoto, S.} \& \bibinfo{author}{Brozović, N.}
\newblock \bibinfo{title}{Aquifer depletion exacerbates agricultural drought losses in the {US} {High} {Plains}}.
\newblock \emph{\bibinfo{journal}{Nature Water}} \textbf{\bibinfo{volume}{2}}, \bibinfo{pages}{41--51} (\bibinfo{year}{2024}).

\bibitem{koningsSensitivityGrasslandProductivity2017}
\bibinfo{author}{Konings, A.~G.}, \bibinfo{author}{Williams, A.~P.} \& \bibinfo{author}{Gentine, P.}
\newblock \bibinfo{title}{Sensitivity of grassland productivity to aridity controlled by stomatal and xylem regulation}.
\newblock \emph{\bibinfo{journal}{Nature Geoscience}} \textbf{\bibinfo{volume}{10}}, \bibinfo{pages}{284--288} (\bibinfo{year}{2017}).

\bibitem{dralleTechnicalNoteAccounting2021}
\bibinfo{author}{Dralle, D.~N.}, \bibinfo{author}{Hahm, W.~J.}, \bibinfo{author}{Chadwick, K.~D.}, \bibinfo{author}{McCormick, E.} \& \bibinfo{author}{Rempe, D.~M.}
\newblock \bibinfo{title}{Technical note: {Accounting} for snow in the estimation of root zone water storage capacity from precipitation and evapotranspiration fluxes}.
\newblock \emph{\bibinfo{journal}{Hydrology and Earth System Sciences}} \textbf{\bibinfo{volume}{25}}, \bibinfo{pages}{2861--2867} (\bibinfo{year}{2021}).

\bibitem{EconomicAnalysis2014}
\bibinfo{author}{{Howitt, R. E.}}, \bibinfo{author}{{Medellín-Azuara, J.}}, \bibinfo{author}{{MacEwan, D.}}, \bibinfo{author}{{Lund, J. R.}} \& \bibinfo{author}{{Sumner, D. A.}}
\newblock \bibinfo{title}{Economic {Analysis} of the 2014 {Drought} for {California} {Agriculture}}.
\newblock \emph{\bibinfo{journal}{UC Davis Center for Watershed Sciences}}  (\bibinfo{year}{2014}).

\bibitem{WaterFutureSan}
\bibinfo{author}{{Hanak, Ellen}} \emph{et~al.}
\newblock \bibinfo{title}{Water and the {Future} of the {San} {Joaquin} {Valley}}.
\newblock \bibinfo{type}{Tech. Rep.}, \bibinfo{institution}{Public Policy Institute of California}, \bibinfo{address}{San Francisco, CA} (\bibinfo{year}{2019}).

\bibitem{humphreySoilMoistureAtmosphere2021}
\bibinfo{author}{Humphrey, V.} \emph{et~al.}
\newblock \bibinfo{title}{Soil moisture–atmosphere feedback dominates land carbon uptake variability}.
\newblock \emph{\bibinfo{journal}{Nature}} \textbf{\bibinfo{volume}{592}}, \bibinfo{pages}{65--69} (\bibinfo{year}{2021}).

\bibitem{fanHydrologicRegulationPlant2017}
\bibinfo{author}{Fan, Y.}, \bibinfo{author}{Miguez-Macho, G.}, \bibinfo{author}{Jobbágy, E.~G.}, \bibinfo{author}{Jackson, R.~B.} \& \bibinfo{author}{Otero-Casal, C.}
\newblock \bibinfo{title}{Hydrologic regulation of plant rooting depth}.
\newblock \emph{\bibinfo{journal}{Proceedings of the National Academy of Sciences}} \textbf{\bibinfo{volume}{114}}, \bibinfo{pages}{10572--10577} (\bibinfo{year}{2017}).

\bibitem{forzieriEmergingSignalsDeclining2022}
\bibinfo{author}{Forzieri, G.}, \bibinfo{author}{Dakos, V.}, \bibinfo{author}{McDowell, N.~G.}, \bibinfo{author}{Ramdane, A.} \& \bibinfo{author}{Cescatti, A.}
\newblock \bibinfo{title}{Emerging signals of declining forest resilience under climate change}.
\newblock \emph{\bibinfo{journal}{Nature}} \textbf{\bibinfo{volume}{608}}, \bibinfo{pages}{534--539} (\bibinfo{year}{2022}).

\bibitem{mcdowellMechanismsWoodyplantMortality2022}
\bibinfo{author}{McDowell, N.~G.} \emph{et~al.}
\newblock \bibinfo{title}{Mechanisms of woody-plant mortality under rising drought, {CO2} and vapour pressure deficit}.
\newblock \emph{\bibinfo{journal}{Nature Reviews Earth \& Environment}} \textbf{\bibinfo{volume}{3}}, \bibinfo{pages}{294--308} (\bibinfo{year}{2022}).

\bibitem{ortiz-bobeaAnthropogenicClimateChange2021}
\bibinfo{author}{Ortiz-Bobea, A.}, \bibinfo{author}{Ault, T.~R.}, \bibinfo{author}{Carrillo, C.~M.}, \bibinfo{author}{Chambers, R.~G.} \& \bibinfo{author}{Lobell, D.~B.}
\newblock \bibinfo{title}{Anthropogenic climate change has slowed global agricultural productivity growth}.
\newblock \emph{\bibinfo{journal}{Nature Climate Change}} \textbf{\bibinfo{volume}{11}}, \bibinfo{pages}{306--312} (\bibinfo{year}{2021}).

\bibitem{gouldenCaliforniaForestDieOff2019}
\bibinfo{author}{Goulden, M.~L.} \& \bibinfo{author}{Bales, R.~C.}
\newblock \bibinfo{title}{California forest die-off linked to multi-year deep soil drying in 2012–2015 drought}.
\newblock \emph{\bibinfo{journal}{Nature Geoscience}} \textbf{\bibinfo{volume}{12}}, \bibinfo{pages}{632--637} (\bibinfo{year}{2019}).

\bibitem{andersonEvaluationDroughtIndices2011}
\bibinfo{author}{Anderson, M.~C.} \emph{et~al.}
\newblock \bibinfo{title}{Evaluation of {Drought} {Indices} {Based} on {Thermal} {Remote} {Sensing} of {Evapotranspiration} over the {Continental} {United} {States}}  (\bibinfo{year}{2011}).

\bibitem{yaoMonitoringDroughtConterminous2010}
\bibinfo{author}{Yao, Y.}, \bibinfo{author}{Liang, S.}, \bibinfo{author}{Qin, Q.} \& \bibinfo{author}{Wang, K.}
\newblock \bibinfo{title}{Monitoring {Drought} over the {Conterminous} {United} {States} {Using} {MODIS} and {NCEP} {Reanalysis}-2 {Data}}  (\bibinfo{year}{2010}).

\bibitem{christianMethodologyFlashDrought2019}
\bibinfo{author}{Christian, J.~I.} \emph{et~al.}
\newblock \bibinfo{title}{A {Methodology} for {Flash} {Drought} {Identification}: {Application} of {Flash} {Drought} {Frequency} across the {United} {States}}  (\bibinfo{year}{2019}).

\bibitem{wongMultiscaleAssessmentAgricultural2021}
\bibinfo{author}{Wong, A.~J.} \emph{et~al.}
\newblock \bibinfo{title}{Multiscale {Assessment} of {Agricultural} {Consumptive} {Water} {Use} in {California}'s {Central} {Valley}}.
\newblock \emph{\bibinfo{journal}{Water Resources Research}} \textbf{\bibinfo{volume}{57}}, \bibinfo{pages}{e2020WR028876} (\bibinfo{year}{2021}).

\bibitem{moyersInvisibleWaterSurcharge2024}
\bibinfo{author}{Moyers, K.}, \bibinfo{author}{Abatzoglou, J.~T.}, \bibinfo{author}{Escriva-Bou, A.}, \bibinfo{author}{Medellín-Azuara, J.} \& \bibinfo{author}{Viers, J.~H.}
\newblock \bibinfo{title}{An invisible water surcharge: {Climate} warming increases crop water demand in the {San} {Joaquin} {Valley}’s groundwater-dependent irrigated agriculture}.
\newblock \emph{\bibinfo{journal}{PLOS Water}} \textbf{\bibinfo{volume}{3}}, \bibinfo{pages}{e0000184} (\bibinfo{year}{2024}).

\bibitem{spinelliWaterStressCauses2016}
\bibinfo{author}{Spinelli, G.~M.}, \bibinfo{author}{Snyder, R.~L.}, \bibinfo{author}{Sanden, B.~L.} \& \bibinfo{author}{Shackel, K.~A.}
\newblock \bibinfo{title}{Water stress causes stomatal closure but does not reduce canopy evapotranspiration in almond}.
\newblock \emph{\bibinfo{journal}{Agricultural Water Management}} \textbf{\bibinfo{volume}{168}}, \bibinfo{pages}{11--22} (\bibinfo{year}{2016}).

\bibitem{fisherFutureEvapotranspirationGlobal2017}
\bibinfo{author}{Fisher, J.~B.} \emph{et~al.}
\newblock \bibinfo{title}{The future of evapotranspiration: {Global} requirements for ecosystem functioning, carbon and climate feedbacks, agricultural management, and water resources}.
\newblock \emph{\bibinfo{journal}{Water Resources Research}} \textbf{\bibinfo{volume}{53}}, \bibinfo{pages}{2618--2626} (\bibinfo{year}{2017}).

\bibitem{fisherEvapotranspirationEverywhereAll2026}
\bibinfo{author}{Fisher, J.~B.} \emph{et~al.}
\newblock \bibinfo{title}{Evapotranspiration {Everywhere}, {All} the {Time}: {Towards} a {Unified} {View} {From} {Earth} {Observation}}.
\newblock \emph{\bibinfo{journal}{Global Change Biology}} \textbf{\bibinfo{volume}{32}}, \bibinfo{pages}{e70898} (\bibinfo{year}{2026}).

\bibitem{schipanskiMovingMeasurementGovernance2023}
\bibinfo{author}{Schipanski, M.~E.} \emph{et~al.}
\newblock \bibinfo{title}{Moving from measurement to governance of shared groundwater resources}.
\newblock \emph{\bibinfo{journal}{Nature Water}} \textbf{\bibinfo{volume}{1}}, \bibinfo{pages}{30--36} (\bibinfo{year}{2023}).

\bibitem{meltonOpenETFillingCritical2022}
\bibinfo{author}{Melton, F.~S.} \emph{et~al.}
\newblock \bibinfo{title}{{OpenET}: {Filling} a {Critical} {Data} {Gap} in {Water} {Management} for the {Western} {United} {States}}.
\newblock \emph{\bibinfo{journal}{JAWRA Journal of the American Water Resources Association}} \textbf{\bibinfo{volume}{58}}, \bibinfo{pages}{971--994} (\bibinfo{year}{2022}).

\bibitem{knipperComparativeAnalysisOpenET2024}
\bibinfo{author}{Knipper, K.} \emph{et~al.}
\newblock \bibinfo{title}{A comparative analysis of {OpenET} for evaluating evapotranspiration in {California} almond orchards}.
\newblock \emph{\bibinfo{journal}{Agricultural and Forest Meteorology}} \textbf{\bibinfo{volume}{355}}, \bibinfo{pages}{110146} (\bibinfo{year}{2024}).

\bibitem{swainIncreasingPrecipitationVolatility2018}
\bibinfo{author}{Swain, D.~L.}, \bibinfo{author}{Langenbrunner, B.}, \bibinfo{author}{Neelin, J.~D.} \& \bibinfo{author}{Hall, A.}
\newblock \bibinfo{title}{Increasing precipitation volatility in twenty-first-century {California}}.
\newblock \emph{\bibinfo{journal}{Nature Climate Change}} \textbf{\bibinfo{volume}{8}}, \bibinfo{pages}{427--433} (\bibinfo{year}{2018}).

\bibitem{NationalLandCover2023}
\bibinfo{author}{{Jon Dewitz}}.
\newblock \bibinfo{title}{National {Land} {Cover} {Database} ({NLCD}) 2019 {Products} (ver. 3.0, {February} 2024)} (\bibinfo{year}{2023}).

\bibitem{YangEtAl2018}
\bibinfo{author}{Yang, L.} \emph{et~al.}
\newblock \bibinfo{title}{A new generation of the {United} {States} {National} {Land} {Cover} {Database}: {Requirements}, research priorities, design, and implementation strategies}.
\newblock \emph{\bibinfo{journal}{ISPRS Journal of Photogrammetry and Remote Sensing}} \textbf{\bibinfo{volume}{146}}, \bibinfo{pages}{108--123} (\bibinfo{year}{2018}).

\bibitem{BoryanEtAl2011}
\bibinfo{author}{Boryan, C.}, \bibinfo{author}{Yang, Z.}, \bibinfo{author}{Mueller, R.} \& \bibinfo{author}{Craig, M.}
\newblock \bibinfo{title}{Monitoring {US} agriculture: the {US} {Department} of {Agriculture}, {National} {Agricultural} {Statistics} {Service}, {Cropland} {Data} {Layer} {Program}}.
\newblock \emph{\bibinfo{journal}{Geocarto International}} \textbf{\bibinfo{volume}{26}}, \bibinfo{pages}{341--358} (\bibinfo{year}{2011}).

\bibitem{SoilSurveyStaffGSSURGO}
\bibinfo{author}{{Soil Survey Staff}}.
\newblock \bibinfo{title}{Gridded {Soil} {Survey} {Geographic} ({gSSURGO}) {Database} for the {United} {States} of {America} and the {Territories}, {Commonwealths}, and {Island} {Nations} served by the {USDA}-{NRCS}}.
\newblock \bibinfo{type}{Database}, \bibinfo{institution}{United States Department of Agriculture, Natural Resources Conservation Service} (\bibinfo{year}{2020}).

\bibitem{3DElevationProgram}
\bibinfo{author}{{William J. Carswell}}.
\newblock \bibinfo{title}{The {3D} {Elevation} {Program}: summary for {California}}.
\newblock \bibinfo{type}{Tech. Rep.} (\bibinfo{year}{2013}).

\bibitem{wang-erlandssonGlobalRootZone2016}
\bibinfo{author}{Wang-Erlandsson, L.} \emph{et~al.}
\newblock \bibinfo{title}{Global root zone storage capacity from satellite-based evaporation}.
\newblock \emph{\bibinfo{journal}{Hydrology and Earth System Sciences}} \textbf{\bibinfo{volume}{20}}, \bibinfo{pages}{1459--1481} (\bibinfo{year}{2016}).

\bibitem{JPLGRACEMascon}
\bibinfo{author}{{Wiese, D. N.}}, \bibinfo{author}{{Yuan, D.-N.}}, \bibinfo{author}{{Boening, C.}}, \bibinfo{author}{{Landerer, F. W.}} \& \bibinfo{author}{{Watkins, M. M.}}
\newblock \bibinfo{title}{{JPL} {GRACE} {Mascon} {Ocean}, {Ice}, and {Hydrology} {Equivalent} {Water} {Height} {Release} 06 {Coastal} {Resolution} {Improvement} ({CRI}) {Filtered} {Version} 1.0}.
\newblock \bibinfo{publisher}{Physical Oceanography Distributed Active Archive Center (PO.DAAC)} (\bibinfo{year}{2018}).

\bibitem{xiaContinentalscaleWaterEnergy2012}
\bibinfo{author}{Xia, Y.} \emph{et~al.}
\newblock \bibinfo{title}{Continental-scale water and energy flux analysis and validation for the {North} {American} {Land} {Data} {Assimilation} {System} project phase 2 ({NLDAS}-2): 1. {Intercomparison} and application of model products}.
\newblock \emph{\bibinfo{journal}{Journal of Geophysical Research: Atmospheres}} \textbf{\bibinfo{volume}{117}} (\bibinfo{year}{2012}).

\bibitem{NOHRSCSNODAS}
\bibinfo{author}{{National Operational Hydrologic Remote Sensing Center}}.
\newblock \bibinfo{title}{Snow data assimilation system ({SNODAS}) data products at {NSIDC}, version 1}.
\newblock \bibinfo{publisher}{NSIDC} (\bibinfo{year}{2004}).

\bibitem{CDEC}
\bibinfo{author}{{California Data Exchange Center}}.
\newblock \bibinfo{title}{Reservoir {Storage}—{Active} {Monthly} {Reservoir} {Report}}.
\newblock \bibinfo{publisher}{California Department of Water Resources} (\bibinfo{year}{2026}).

\end{thebibliography}

\clearpage
\appendix

\clearpage
\appendix

\setcounter{figure}{0}
\renewcommand{\thefigure}{S\arabic{figure}}
\setcounter{table}{0}
\renewcommand{\thetable}{S\arabic{table}}

\begin{center}
  {\Large\bfseries Supplementary Information\par}
  \vspace{1em}
\end{center}

\vspace{2em}

\section*{Supplementary figures}

\begin{figure}[H]
\centering
\includegraphics[width=1.05\textwidth]{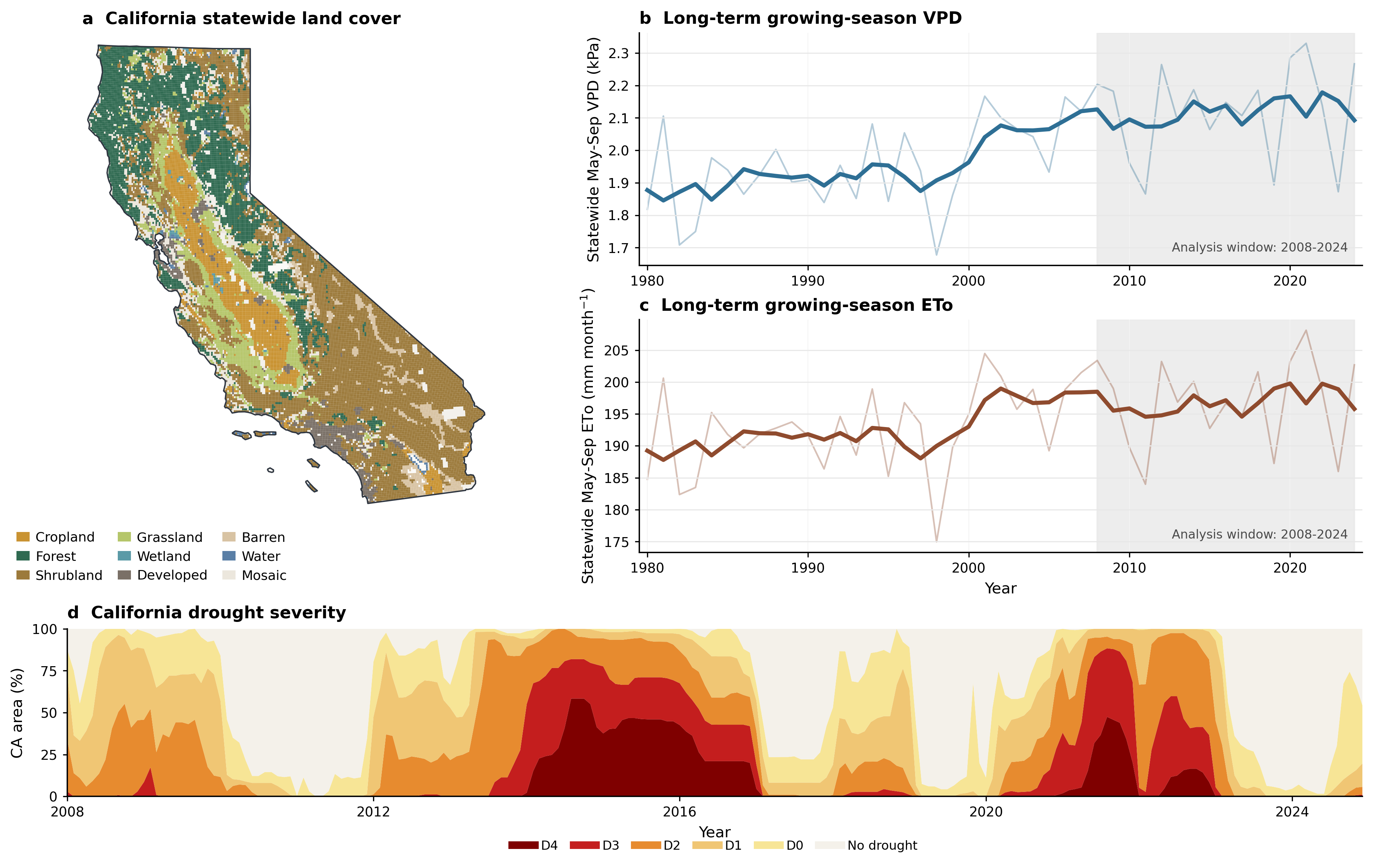}
\caption{\textbf{California land-cover diversity, atmospheric evaporative demand, and drought exposure.} (a) Statewide land-cover distribution across California, showing forests, grasslands, shrublands, croplands, developed areas, barren land, wetlands, and water. (b,c) Long-term statewide May-September vapour pressure deficit (VPD); and reference evapotranspiration (ETo) from 1980 to 2024. Thin lines show annual growing-season means, thick lines show five-year moving averages, and shading marks the 2008-2024 analysis window. (d) Fraction of California area classified under each U.S. Drought Monitor category during the analysis period.}
\end{figure}

\begin{figure}[H]
\centering
\includegraphics[width=1.05\textwidth]{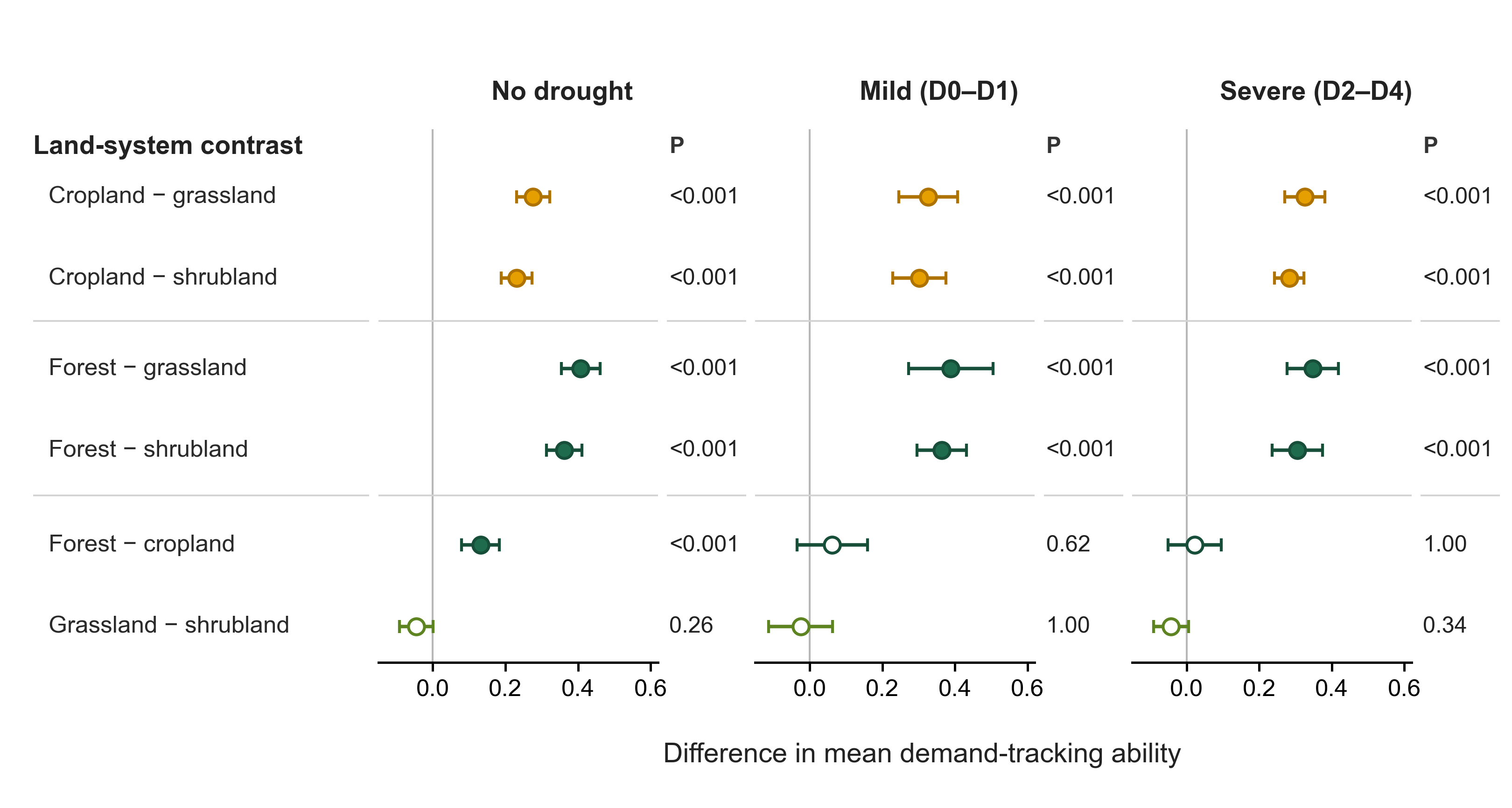}
\caption{\textbf{Land-system contrasts in demand-tracking ability across drought classes.} Pairwise differences in mean demand-tracking ability were estimated within each drought class using a linear model including land system, drought class and their interaction. Points show estimated differences and horizontal bars show 95\% confidence intervals. Positive values indicate higher demand-tracking ability in the first-listed land system. Confidence intervals and P values account for spatial dependence using 100-km spatial blocks. Filled points indicate P$<0.05$ after Holm correction for multiple comparisons.}
\end{figure}

\begin{figure}[H]
\centering
\includegraphics[width=1.05\textwidth]{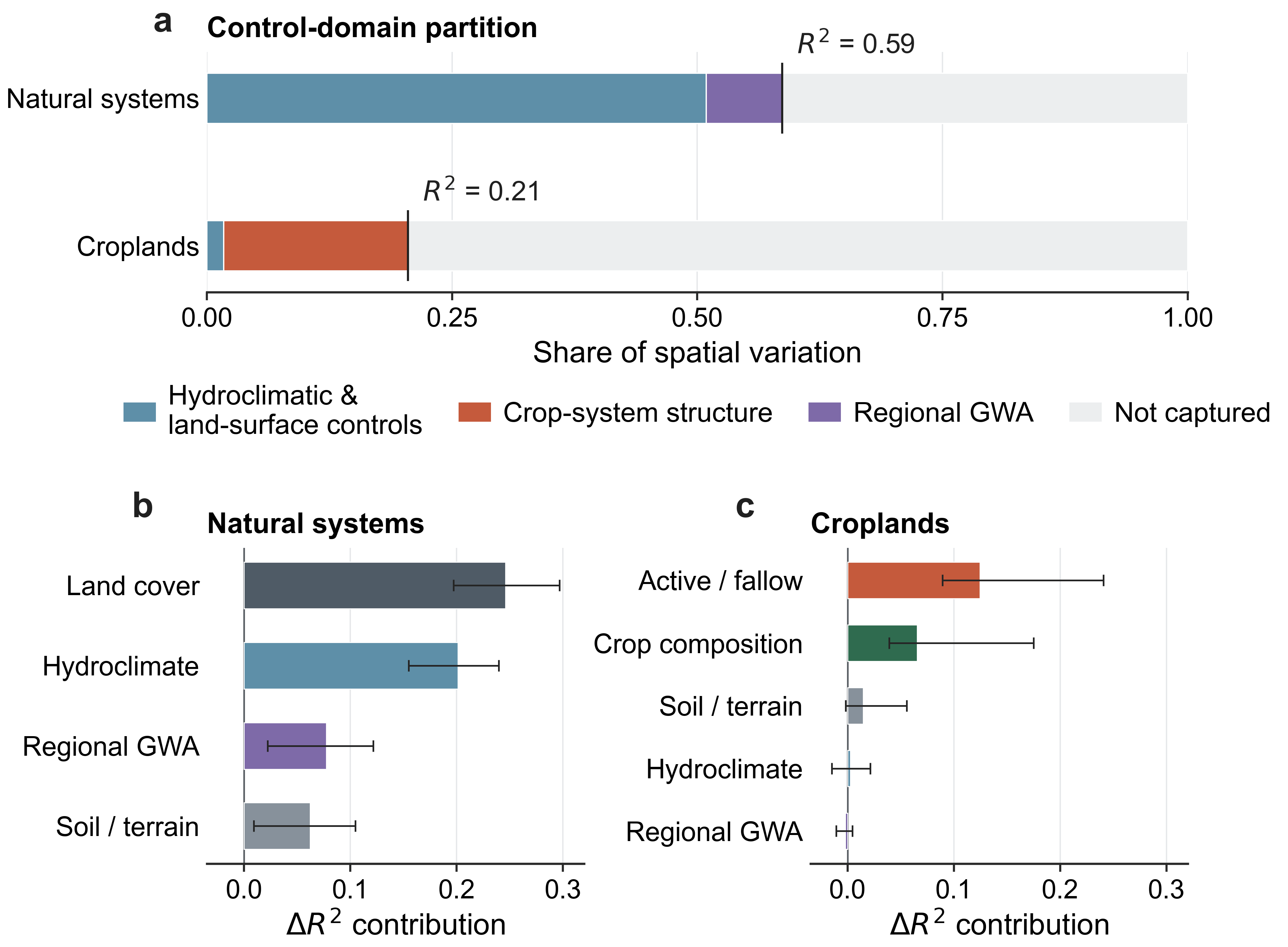}
\caption{\textbf{Static information domains organize demand tracking differently in natural systems and croplands.} (a) Diagnostic partition of spatial variation in mean demand-tracking ability captured by static information domains. Coloured segments show order-averaged, spatial cross-validated $\Delta R^2$ contributions, and grey segments indicate variation not captured by these domains. Natural-system domains include land cover, hydroclimate, soil/terrain and regional groundwater-storage anomaly (GWA). Cropland domains include hydroclimate, soil/terrain, active/fallow state, crop composition and regional GWA. (b,c) Domain contributions for natural systems and croplands. Error bars show 95\% spatial-block bootstrap intervals. For croplands, the available domains characterize crop-system structure but do not include annual surface-water deliveries, groundwater pumping, irrigation allocations, or field-level irrigation practices. The partition is interpreted as a partial diagnostic of spatial information, not as causal attribution or a comprehensive explanation of water access.}
\end{figure}

\begin{figure}[H]
\centering
\includegraphics[width=1.05\textwidth]{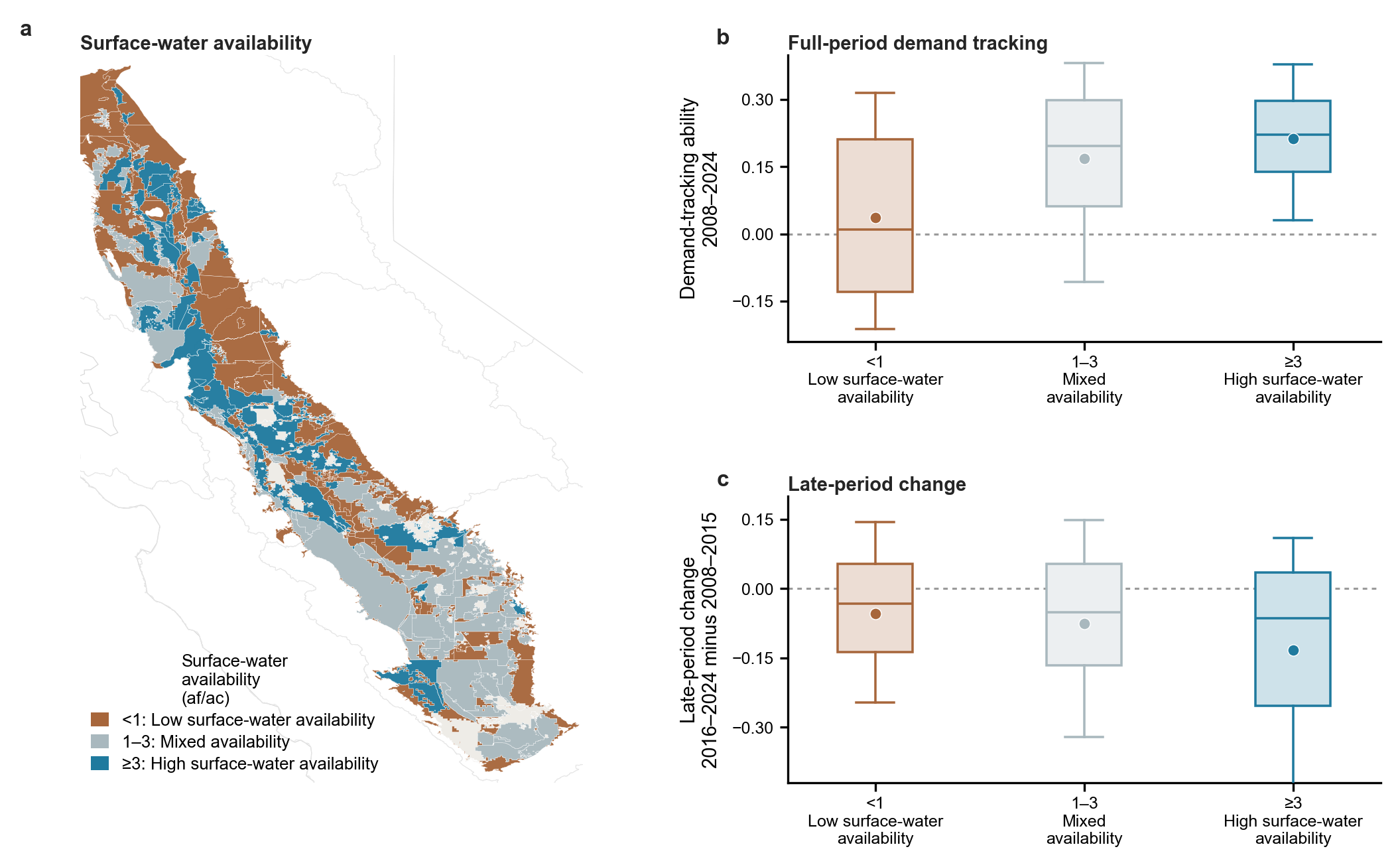}
\caption{\textbf{Surface-water availability and cropland demand tracking in the Central Valley.} (a) Surface-water availability classes across Central Valley water-management entities. Surface-water availability is defined from Public Policy Institute of California datasets as historical average surface-water deliveries per irrigated acre and grouped as low ($<1$ af ac$^{-1}$), mixed (1-3 af ac$^{-1}$) and high ($\geq3$ af ac$^{-1}$). (b) Full-period cropland demand-tracking ability from 2008 to 2024 by surface-water availability class. (c) Late-period change in cropland demand-tracking ability, calculated as 2016-2024 minus 2008-2015. Negative values indicate weakened demand tracking in the late period. The PPIC surface-water availability metric is used as a static water-portfolio proxy, not as annual surface-water use, groundwater pumping, or irrigation allocation.}
\end{figure}

\begin{figure}[H]
\centering
\includegraphics[width=1.05\textwidth]{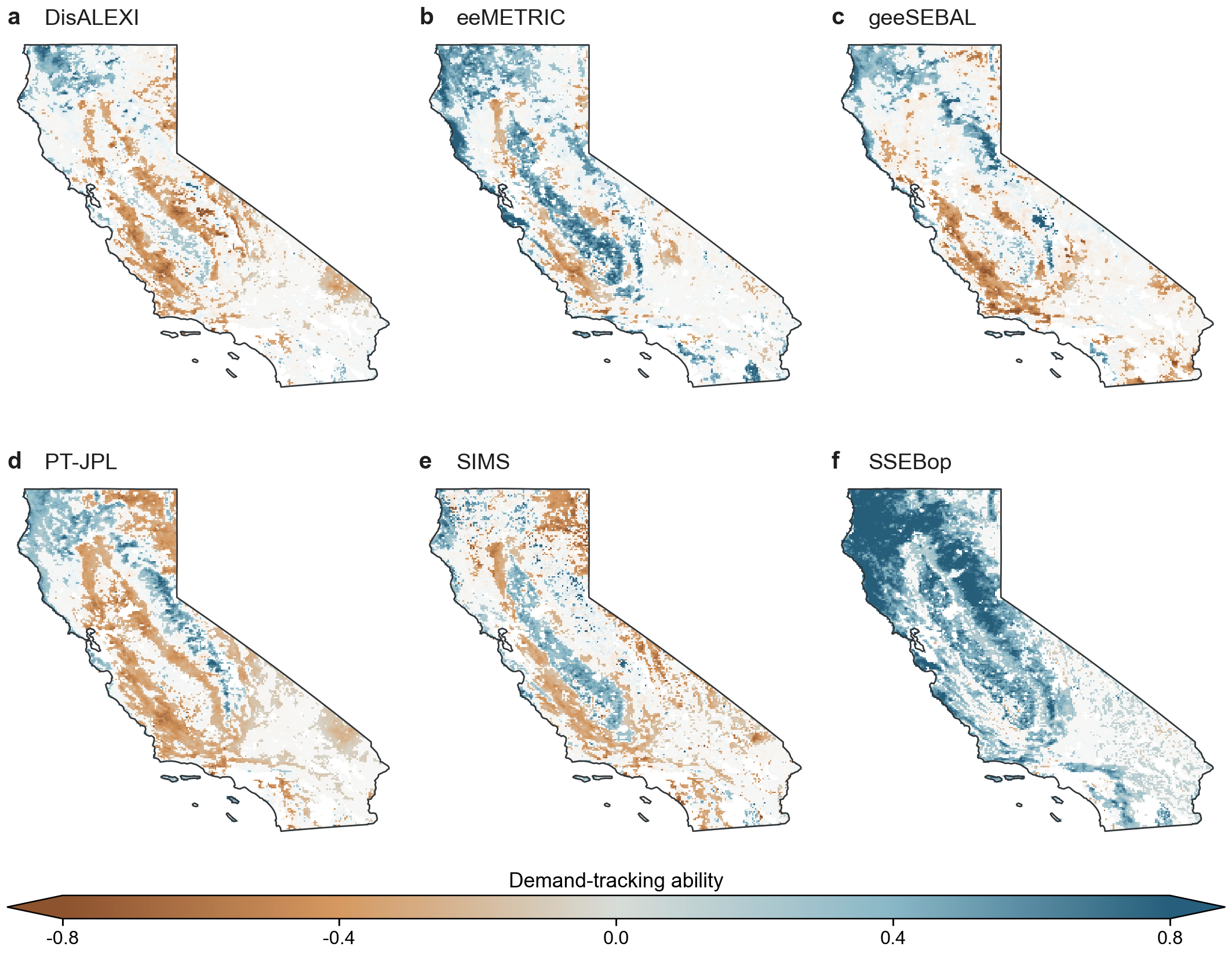}
\caption{\textbf{Demand-tracking ability estimated from individual OpenET component models.} (a-f) Statewide demand-tracking ability calculated separately using ETa from each OpenET component model: DisALEXI (a), eeMETRIC (b), geeSEBAL (c), PT-JPL (d), SIMS (e) and SSEBop (f). For each model, demand-tracking ability was estimated using the same May–September ETa–ETo anomaly-slope definition and the same 4-km spatial support as the main ensemble-based analysis. Blue indicates positive ETa–ETo anomaly coupling, while brown indicates weak or negative coupling. The component-model maps show that the broad contrast between positively tracking forest or irrigated regions and weakly tracking dry interior landscapes is not unique to a single OpenET model, although individual models differ in magnitude and spatial detail.}
\end{figure}

\begin{figure}[H]
\centering
\includegraphics[width=1.05\textwidth]{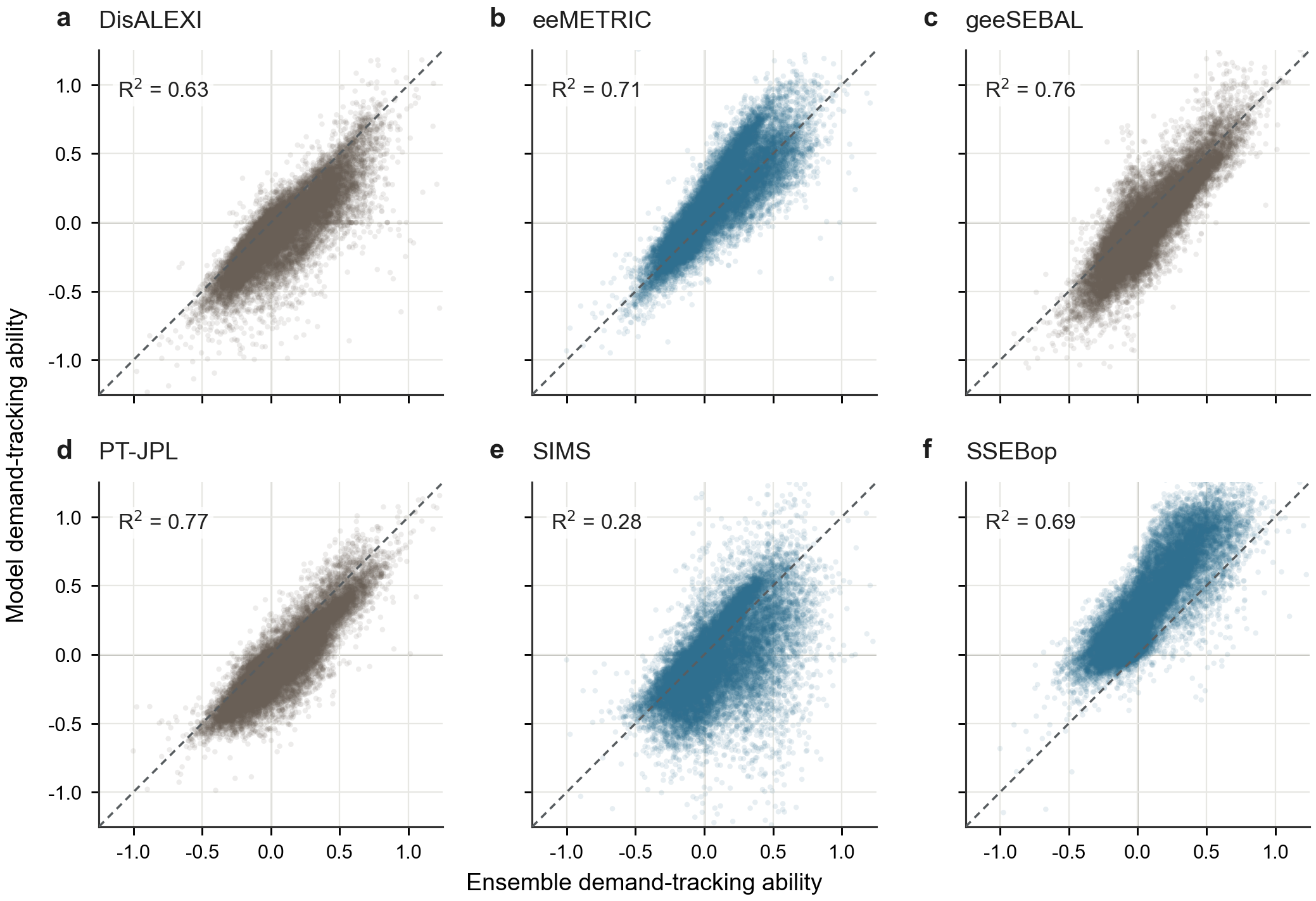}
\caption{\textbf{Agreement between component-model and ensemble demand-tracking ability.} (a-f) Scatterplots comparing demand-tracking ability estimated from each OpenET component model against the OpenET ensemble estimate across common valid grid cells. Dashed lines show the one-to-one relationship. Reported $R^2$ values summarize linear agreement between each component model and the ensemble: DisALEXI, $R^2$=0.63; eeMETRIC, $R^2$=0.71; geeSEBAL, $R^2$=0.76; PT-JPL, $R^2$=0.77; SIMS, $R^2$=0.28; and SSEBop, $R^2$=0.69. Most component models show moderate to strong agreement with the ensemble, whereas SIMS shows weaker agreement, consistent with its distinct crop-coefficient formulation and stronger dependence on crop type and canopy-density assumptions.}
\end{figure}

\begin{figure}[H]
\centering
\includegraphics[width=1.05\textwidth]{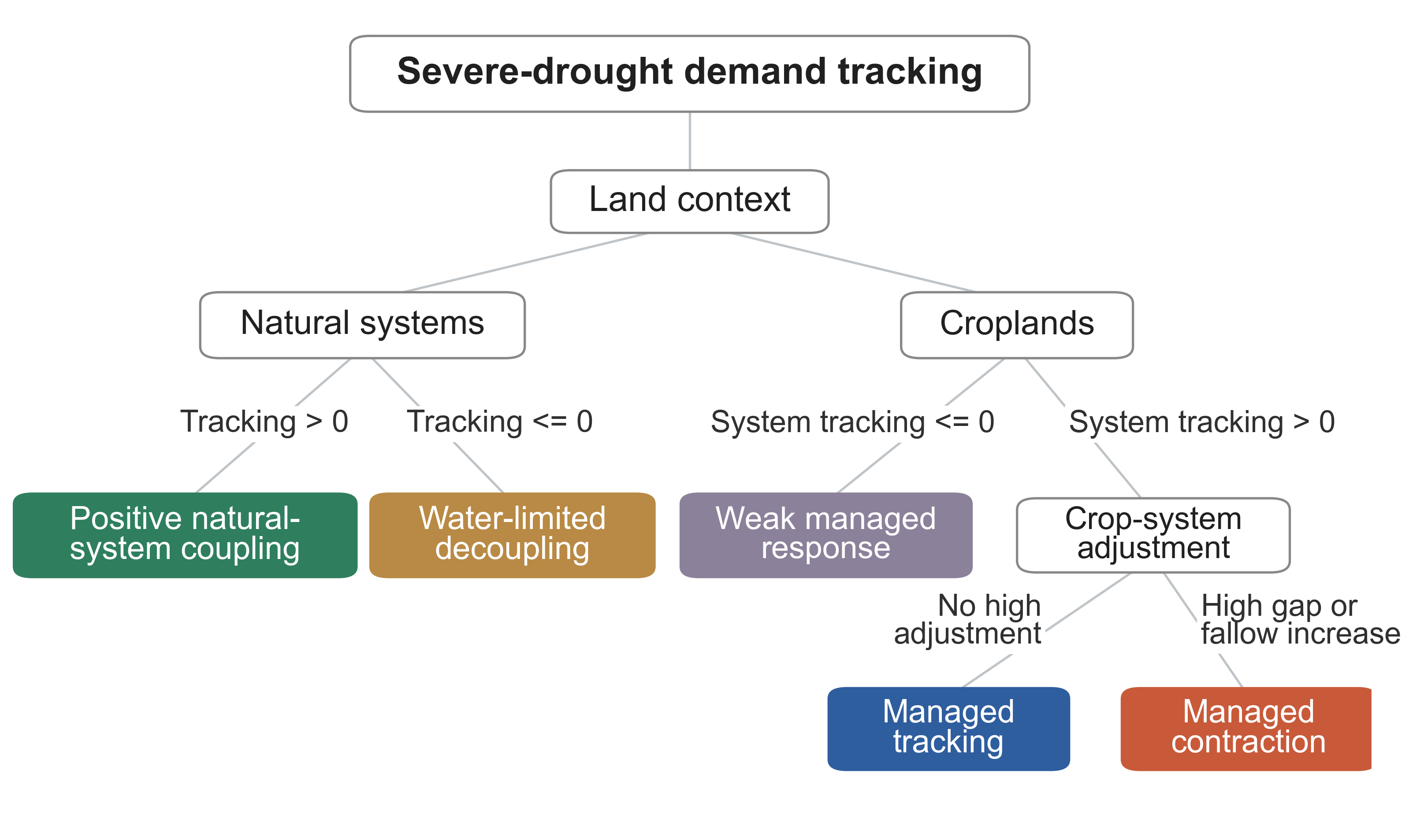}
\caption{\textbf{Classification workflow for severe-drought response states.} Classification rules used to assign severe-drought response states from valid demand-tracking ability estimates. Natural systems were classified as positive natural-system coupling or water-limited decoupling according to whether severe-drought demand-tracking ability was positive or non-positive. Croplands were classified as weak managed response, managed tracking, or managed contraction using whole crop-system demand-tracking ability, the active-field versus whole crop-system tracking gap, and the increase in crop-attributed fallow share.}
\end{figure}

\section*{Supplementary methods}

\subsection*{OpenET component-model sensitivity}

To evaluate whether demand-tracking patterns were specific to the OpenET ensemble product, we repeated the growing-season anomaly-slope calculation for the six OpenET component models: DisALEXI, eeMETRIC, geeSEBAL, PT-JPL, SIMS and SSEBop. For each component model, monthly ETa was reconstructed on the same 4-km grid from active, fallow and natural ET components weighted by their corresponding land fractions. We used the same 2008-2024 May-September window and calculated ETa and ETo anomalies relative to each grid cell and calendar month. Cell-level component-model slopes were retained when sufficient observations and ETo-anomaly variability were available, and were compared with the ensemble-based slopes on matched grid cells. These analyses were used only as product-sensitivity checks and were not used to define the primary response states.

\subsection*{Spatial information partitioning} 

We quantified how static information domains organized spatial variation in full-period demand-tracking ability separately for natural systems and croplands. The response variable was the 2008-2024 May-September demand-tracking slope for each valid 4-km analysis unit. Natural-system domains included land cover, hydroclimate, soil/terrain and regional GWA. Cropland domains included hydroclimate, soil/terrain, regional GWA, active/fallow structure and crop composition. Hydroclimate included growing-season mean vapour pressure deficit, precipitation and temperature; soil/terrain included soil available water capacity, elevation and slope; active/fallow structure included irrigated and fallow area fractions; and crop composition included annual, perennial and major crop-subgroup fractions.

Continuous predictors were standardized within each analysis subset after complete-case filtering. Each predictor was z-standardized by subtracting its subset mean and dividing by its subset standard deviation, resulting in a mean of zero and a standard deviation of one within that subset. Categorical predictors were one-hot encoded with one reference level omitted. Predictors with zero variance or exact collinearity were removed before model fitting. We fitted ordinary least-squares models with an intercept and used spatial-block cross-validation to estimate spatially generalizable variation rather than in-sample fit. Domain contributions were calculated as order-averaged increments in spatial cross-validated \(R^2\), refitting nested models across all possible domain orderings. The natural-system partition used all 24 orderings of four domains, and the cropland partition used all 120 orderings of five domains. Uncertainty intervals were estimated using 1,000 spatial-block bootstrap replicates. These partitions were interpreted as diagnostic associations between static information domains and demand-tracking patterns, not as causal attribution of local water access.

\subsection*{Central Valley water-portfolio context}

Surface-water availability was derived from Public Policy Institute of California (PPIC) water-management-entity datasets, which report historical average surface-water deliveries per irrigated acre for irrigation districts, water districts, groundwater sustainability agency (GSA) member entities, and county-administered areas. The San Joaquin Valley dataset reports average deliveries for water years 2001–2015, while the Sacramento Valley and Delta dataset reports average deliveries for 2010–2019 for most districts. Irrigated acreage and land-use summaries are based on California Department of Water Resources (DWR) 2016 land-use data. We used the PPIC acre-feet per irrigated acre metric as a static proxy for surface-water availability, not as an annual measure of actual surface-water use or groundwater pumping.

\end{document}